\documentclass[prl, superscriptaddress, notitlepage, twocolumn, nofootinbib]{revtex4-1}
\usepackage{amsmath,amsthm,amsfonts,amssymb}
\usepackage{eucal}
\usepackage{tensor}
\usepackage{graphicx}
\usepackage{tabu}
\usepackage{indentfirst}
\usepackage{bbm}
\usepackage{grffile}
\usepackage[all,knot]{xy}
\usepackage{chngcntr}
\usepackage{floatrow}
\usepackage[caption=false]{subfig}
\usepackage[colorlinks, citecolor=blue, linkcolor=blue]{hyperref}
\usepackage[usenames, dvipsnames]{color}
\usepackage{enumitem,kantlipsum}
\usepackage{amsbsy}
\usepackage{multirow}
\usepackage{mathrsfs}
\xyoption{arc}

\newcommand{\A}{\mathcal{A}}

\newcommand{\CC}{\mathcal{C}}

\newcommand{\SUM}{\text{SUM}}

\DeclareMathOperator{\Span}{Span}

\newcommand{\one}{\mathbf{1}}
\newcommand{\C}{\mathbb C}
\newcommand{\mC}{\mathcal{C}}

\newcommand{\mfD}{\mathfrak{D}}
\newcommand{\Z}{\mathbb Z}

\newcommand{\B}{\mathcal{B}}

\newcommand{\comments}[1]{}

\newcommand{\ket}[1]{|#1\rangle}

\renewcommand{\one}{\mathbf{1}}

\renewcommand{\CC}{\mathcal{C}}

\renewcommand{\Z}{\mathbb{Z}}

\newcommand{\overbar}[1]{\mkern 2.3mu\overline{\mkern-2.3mu#1\mkern-2.3mu}\mkern 2.3mu}

\theoremstyle{definition}

\begin{document}

\title{Universal Quantum Computation with Gapped Boundaries}
\author{Iris Cong}
\affiliation{Department of Computer Science, University of California, Los Angeles, CA 90095, U.S.A.} 
\affiliation{Department of Physics, Harvard University, Cambridge, MA 02138, U.S.A.}
\affiliation{Microsoft Station Q, University of California, Santa Barbara, CA 93106-6105 U.S.A.}
\author{Meng Cheng}
\affiliation{Department of Physics, Yale University, New Haven, CT 06520-8120, U.S.A.} 
\affiliation{Microsoft Station Q, University of California, Santa Barbara, CA 93106-6105 U.S.A.}
\author{Zhenghan Wang}
\affiliation{Microsoft Station Q, University of California, Santa Barbara, CA 93106-6105 U.S.A.}
\affiliation{Department of Mathematics, University of California, Santa Barbara, CA 93106-6105 U.S.A.}

\begin{abstract}
This Letter discusses topological quantum computation with gapped boundaries of two-dimensional topological phases. Systematic methods are presented to encode quantum information topologically using gapped boundaries, and to perform topologically protected operations on this encoding. In particular, we introduce a new and general computational primitive of {\it topological charge measurement} and present a symmetry-protected implementation of this primitive. Throughout the Letter, a concrete physical example, the $\Z_3$ toric code ($\mfD(\Z_3)$), is discussed. For this example, we have a qutrit encoding and an abstract universal gate set. Physically, gapped boundaries of $\mfD(\Z_3)$ can be realized in bilayer fractional quantum Hall $1/3$ systems. If a practical implementation is found for the required topological charge measurement, these boundaries will give rise to a direct physical realization of a universal quantum computer based on a purely abelian topological phase.
\end{abstract}

\maketitle

\noindent
{\bf Introduction.} The quantum model of computation strikes a delicate balance between classical digital and analog computing models, as its stability lies closer to digital models, while its computational power is closer to analog ones.  Still, a major obstacle to developing quantum computers lies in the susceptibility of qubits to decoherence. One elegant theoretical solution to this problem is topological quantum computation (TQC) \cite{Free98, Kitaev97, FKLW}. TQC is a paradigm that information is encoded in topological degrees of freedom of certain quantum systems, thereby protected from local decoherence. While the standard implementation uses (non-abelian) anyons in topological phases of matter, recent studies revealed that certain topological phases also support gapped boundaries. It is hence natural to study TQC with gapped boundaries \cite{Bravyi98,Cong16a,Barkeshli16,Fowler12}.

Real samples of topological phases of matter such as fractional quantum Hall liquids and topological insulators have boundaries, which are usually conducting  (gapless) even though the bulk are insulating (gapped).  However, they can be modified to realize Dijkgraaf-Witten (DW) gauge theories, which are also given by Kitaev's quantum double Hamiltonian \cite{Kitaev97}. These theories support gapped boundaries in the sense that the extensions of the Hamiltonians to spaces (surfaces) with boundaries are still gapped; the Hamiltonian and algebraic frameworks are developed in Refs. \cite{Cong16a,Cong16b}.  These frameworks show that a gapped boundary effectively behaves as a non-abelian anyon. However, while the existence of non-abelian anyons is still uncertain, gapped boundaries of abelian phases are much more routine 
and support topologically protected degeneracies even on the plane.

In this Letter, we apply our theory to a concrete physical example---the $\Z_3$ toric code $\mfD(\Z_3)$---to obtain a universal gate set, which is a striking example of the extra computational power from gapped boundaries. This new direction opens up new vistas in both the theoretical study and experimental realization of TQC. We introduce a new computational primitive---topological charge measurement (TCM), which extends topological charge projection \cite{Barkeshli16}.  We propose a physical realization of symmetry-protected TCM in a fractional quantum spin Hall state, while leaving a fully topologically protected one to the future because which measurement is possible in gauge theory is an open fundamental question \cite{Beckman02}.

Our universal gate set for $\mfD(\Z_3)$ is close to experimental technology in bilayer quantum Hall liquids. 
If a practical implementation is found for our TCM primitive, this gate set is a direct physical realization of a universal quantum computer.

\vspace{1mm}
\noindent
{\bf Realization of $\Z_3$ toric code by bilayer $\nu=1/3$ fractional quantum Hall liquids.} The $\Z_3$ toric code $\mfD(\Z_3)$ can be realized in bilayer fractional quantum Hall systems: Ref. \cite{Bark16} considers an electron-hole bilayer FQH system, with a $1/3$ Laughlin state of opposite chirality in each layer. The topological order in this system is $\mathrm{SU}(3)_1\times\overline{\mathrm{SU}(3)_1}$~\footnote{Together with physical electrons, $\overline{\mathrm{SU}(3)_1}$ is topologically equivalent to a $1/3$ Laughlin state}, which is equivalent to the $\Z_3$ toric code $\mfD(\Z_3)$.  Hence, we will recycle many of the results of Ref. \cite{Bark16}.


We briefly summarize the basic data for $\mfD(\Z_3)$. Mathematically, a topological phase is described by a modular tensor category (MTC) $\B$ \cite{BakalovKirillov}. 
The anyon types\footnote{There are many terms in the literature referring to the same thing: a simple quasiparticle, an anyon, or a simple object of $\mfD(G)$. An anyon type, topological charge, or superselection sector is an isomorphism class of the above.} are $e^a m^b$, $a,b = 0,1,2$, where $e$ and $m$ are $\mathbb{Z}_3$ unit gauge charge and flux respectively (so $e^2 = \overbar{e}$, $m^2 = \overbar{m}$). 
The braiding statistics of the anyons is encoded in
the modular $\mathcal{S} = [S_{ab}]$ and $\mathcal{T} = [T_{ab}]$ matrices \cite{BakalovKirillov}:
\begin{equation}
S_{ab} = \omega^{-a_2 b_1 - a_1 b_2},
\qquad
T_{ab} = \omega^{a_1a_2}\delta_{ab}.
\end{equation}
Here, $\omega = e^{2 \pi i /3}$.

\vspace{1mm}
\noindent
{\bf Gapped boundaries, degeneracy, and topological operations.} Let us first review the physics of gapped boundaries, as they will encode our topological qudits. A convenient physical description for a gapped boundary type is the consistent collection of (bosonic) anyons that can condense to vacuum to the boundary at no energy cost. Mathematically, this is given by a Lagrangian algebra $\mathcal{A}$ in the MTC $\B$\footnote{See Refs. \cite{Cong16a,Cong16b,KitaevKong,Beigi11,LWW} and references therein for precise definitions.}, which can be represented as a direct sum of all condensed anyon types.
For $\mfD(\Z_3)$, there are two gapped boundary types: $e$-boundary (resp. $m$-boundary) where $e,e^2$ (resp. $m, m^2$) condense. The corresponding Lagrangian algebras are $1+e +e^2$ and $1 +m +m^2$. In the bilayer $1/3$ Laughlin state description, the $e$/$m$-boundary types correspond to holes with the two layers connected via electron pairing (i.e. superconducting) or tunneling.



\begin{figure}
\centering
\includegraphics[width = 0.58\textwidth]{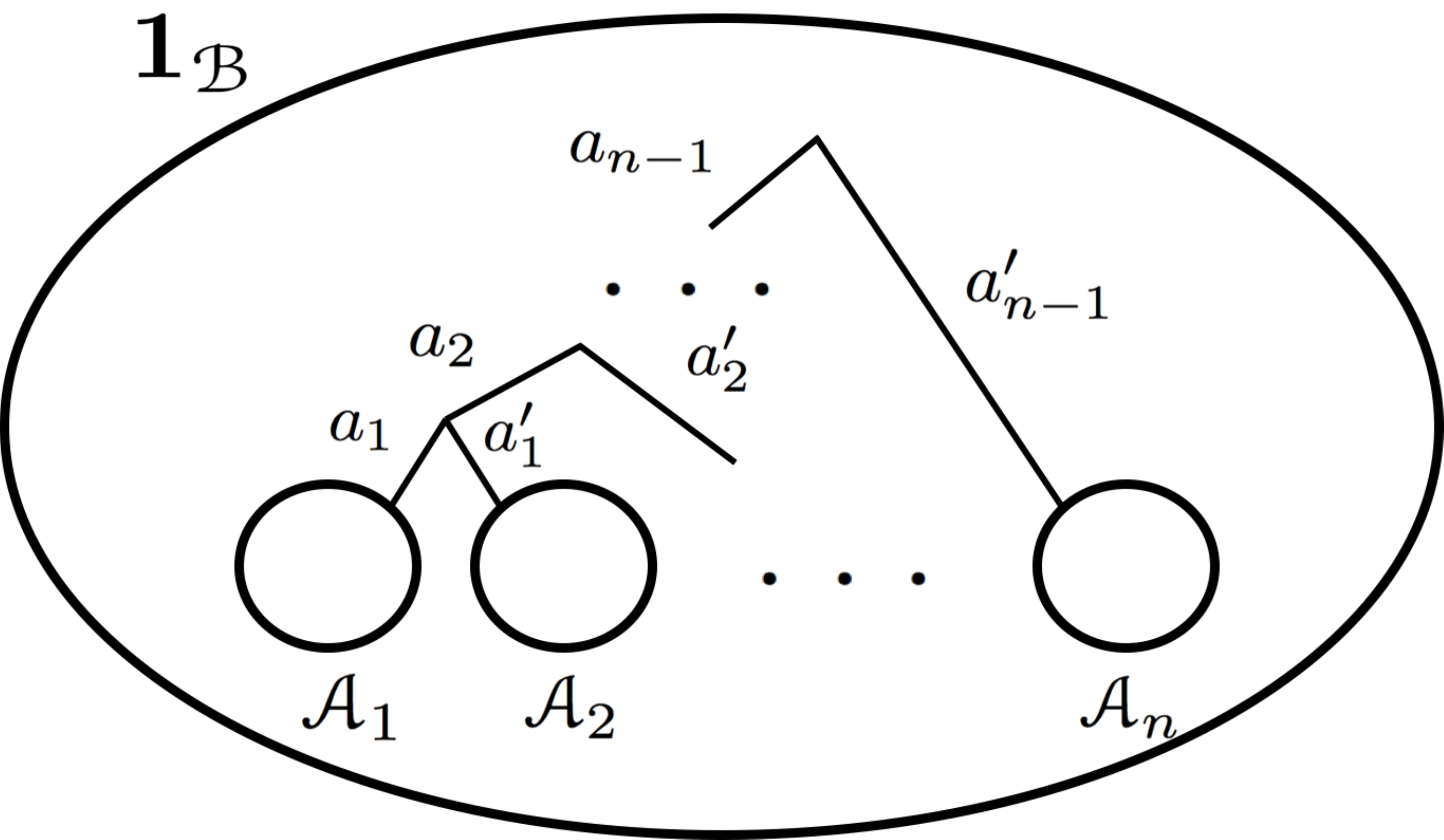}
\caption{Ground state for $n$ gapped boundaries $\A_i$ on a plane and total charge vacuum. All edges are directed to point downward.}
\label{fig:algebraic-gsd-n-2}
\end{figure}

Multiple gapped boundaries support a degenerate ground state manifold~\footnote{The degeneracy is exponentially protected in all length scales, including distance between boundaries as well as lengths of the boundaries.}.  Consider a closed system with $n$ gapped boundaries (Fig. \ref{fig:algebraic-gsd-n-2}). Refs. \cite{Cong16a,Cong16b} show that the ground state of the system is given by the different ways we can create $n$ anyons out of vacuum, and condense all of them onto the boundaries as a fusion tree (Fig. \ref{fig:algebraic-gsd-n-2}). This fusion tree also specifies a choice of basis states for the ground state manifold. For example, if we have two $e$-boundaries in a planar $\mfD(\Z_3)$ theory, the GSD is $3$, labeled by $a_1=\bar{c}$, $a_2=c$, $c=\one,e,\bar{e}$. We denote the basis elements by $\ket{c}$ and encode our qutrit in this space.

We now discuss the topological operations on gapped boundaries, which induce unitary transformations in the degenerate subspace. We focus on the $\mfD(\Z_3)$ example and leave the general results to the appendix.

\vspace{1mm}
\noindent
{\it Tunnel-$a$ operations}. The first topological operation is to tunnel an anyon $a$ from one gapped boundary ($\A_1$) to another ($\A_2$), where $a$ (resp. $\overbar{a}$) condenses on $\A_1$ (resp. $\A_2$). Physically, this corresponds to applying the $a$ string operator \cite{Kitaev97} along a path $\gamma$  connecting the two gapped boundaries. This operation, known as a {\it Wilson line operator}, is denoted by $W_a(\gamma)$. For the $\mfD(\Z_3)$ theory, it can be represented as follows:
\begin{equation}
\label{eq:tunnel-formula}
W_a(\gamma) \ket{b}
= \ket{a \times b}.
\end{equation}
Expressing $W_a(\gamma)$ as a matrix that acts on the ground state subspace, we see that $W_e (\gamma)$ implements the single-qutrit Pauli-X gate $\sigma^x_3$.

\vspace{0.5mm}
\noindent
{\it Loop-$a$ operations.} Analogously, one can create a pair of anyons $a,\overbar{a}$ in the bulk, loop one of them around a gapped boundary, and annihilate the pair. When we loop $a$ counter-clockwise around the boundary, this is known as the {\it Wilson loop operator} $W_a (\alpha_i)$ where $\alpha_i$ is the loop encircling boundary $\A_i$. Appendix B shows that
\begin{equation}
\label{eq:loop-formula}
W_a(\alpha_2) \ket{b} = \frac{S_{ab}}{d_b} \ket{b}.
\end{equation}

\begin{figure}
\centering
\includegraphics[width = 0.4\textwidth]{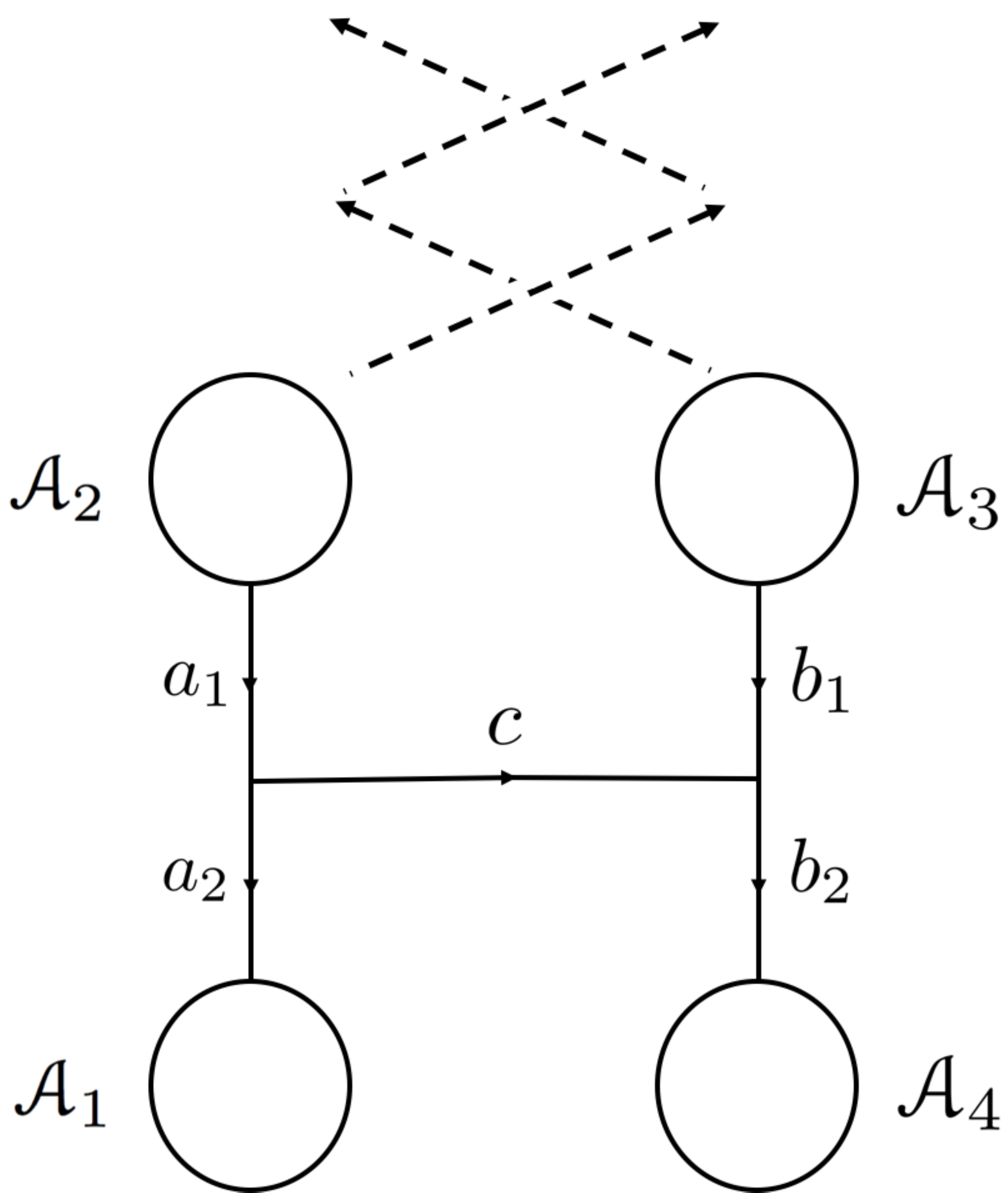}
\caption{Braiding of two gapped boundaries ($\sigma_2^2$). Solid lines indicate tunneling operators from the basis vectors (i.e. {\it not} motion of the holes), while dotted lines indicate how the holes move in the braiding process.}
\label{fig:braiding}
\end{figure}

\noindent
{\it Braiding gapped boundaries.}
Another topological operation is to braid gapped boundaries around each other. This gives multiple-qudit operations that can produce entangling gates. Physically, braiding corresponds to moving gapped boundaries around each other, e.g. by tuning the Hamiltonian $H_{\text{G.B.}}$ of Refs. \cite{Cong16a,Cong16b} adiabatically.

 We may arbitrarily braid $n$ gapped boundaries with total charge vacuum around each other to obtain a unitary transformation on the ground state,  so long as we return each boundary to its original position. Mathematically, this means that the braiding matrices form a representation of the (spherical) $n$-strand pure braid group $P_n$ \cite{Escobar17}. They can be computed using the diagrammatic rules of anyon models and the basis states of gapped boundaries.  For most purposes of quantum computation, it is sufficient to consider 2-qudit encodings, where $n = 4$ (Fig. \ref{fig:braiding}). In general, one must compute all 6 generators of $P_4$. As an example, we derive the formula for the generator $\sigma_2^2$ in Appendix B.

\begin{figure}
\centering
\includegraphics[width = 0.95\textwidth]{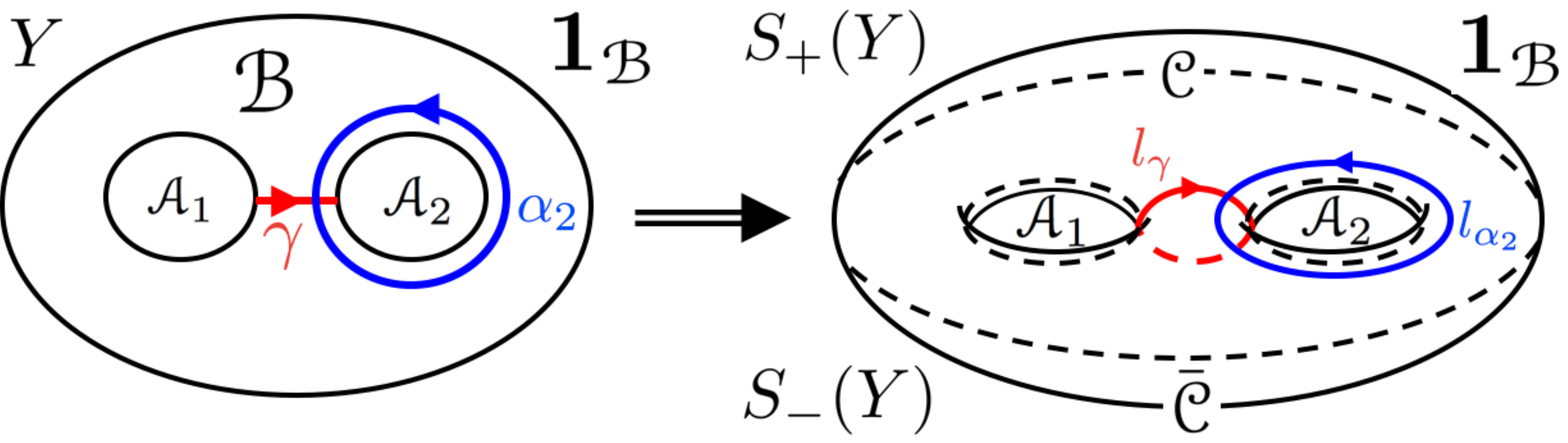}
\caption{Topological charge projection ($n=2$).}
\label{fig:tcm}
\end{figure}

\vspace{1mm}
\noindent
\textbf{Topological charge measurement.}
For a DW theory, the gapped boundary braidings only generate a finite group \cite{Escobar17}. Inspired by the results of Ref. \cite{Barkeshli16}, we introduce topological charge measurement based on the Wilson operators. Before we discuss the general case, recall that topological charge projection can detect the total charge of a collection of quasiparticles inside a certain region, by e.g. sending a probe particle along a path enclosing the region and performing interferometric measurement. As a generalization, we can use similar methods to perform measurement of topological charge through any loop, not just contractible ones, possibly on a higher-genus surface \cite{Barkeshli16}.

Recall that $\mfD(\Z_3)$ splits into two theories $\B = \mC \boxtimes \overbar{\mC}$  with $\mC=\mathrm{SU}(3)_1$ which do not interact in the bulk, but are ``stuck together'' at the original boundaries of $\B$. The planar region $Y$ also splits into two mirror layers, $S_{+}(Y)$ and $S_{-}(Y)$, which are completely disjoint in the bulk but ``stuck together'' at the boundaries of $Y$. This way, we can view the system as a single layer of $\mC$ on a higher-genus surface. 
Similarly, each loop $\alpha$ in $Y$ becomes a loop $l_{\alpha}$ in $S_{+}(Y)$ or $S_{-}(Y)$, while an arc $\gamma$ connecting two boundaries lifts to a loop $l_{\gamma}$ going around both layers. Let $\beta$ be one of these loops. Fig. \ref{fig:tcm} illustrates this for $n=2$.

Define $\mathcal{O}_x(\beta)=W_x(\alpha_i)$ (tunneling operator in $\mC$) if $\beta$ is the lifting of the line $\alpha_i$, and $\mathcal{O}_x(\beta)=W_{x\overbar{x}}(\gamma_i)$ (loop operator in $\B$) if $\beta$ is the lifting of the loop $\gamma_i$. By Ref. \cite{Barkeshli16}, the projection measuring topological charge $a$ through $\beta$ can be expressed as 
\begin{equation}
\label{eq:tcp}
P^{(a)}_{\beta}=\sum_{x\in \CC}S_{0a}S_{xa}^{*} \mathcal{O}_x(\beta). 
\end{equation}
\noindent
The sum runs over the anyon labels $x$ of $\mC$, and $S_{ab}$ is the modular $\mathcal{S}$-matrix of $\CC$. The Wilson operators $W_x(\alpha_i)$ and $W_{x\overbar{x}}(\gamma_i)$ are computed using the formulas (\ref{eq:tunnel-formula}) and (\ref{eq:loop-formula}) with the data of $\mC$ and $\B$, respectively. As shown in \cite{Barkeshli16}, topological charge projections generate all mapping class group representations $V_{\CC}(Y)$ of a closed surface $Y$ from the anyon theory $\CC$.

For our purpose, we generalize these projections to topological charge measurements (TCM) which are the complements of topological charge projections (the more general definition is in Appendix C). Specifically, given an anyon label $a$ and the lifting $\beta$ of a Wilson line/loop as above, we consider the projection $1-P^{(a)}_{\beta}$. Physically, this can be implemented by adding such non-local operators to the effective Hamiltonian of the ground state subspace: 
\begin{equation}
	H'=-t W_a (\beta) +\text{h.c.}
	\label{}
\end{equation}
Here, $t$ is the (complex) tunneling amplitude. This effective Hamiltonian then projects the system to the desired state space.




\vspace{1mm}
\noindent
{\bf Universal gate set with $\mfD(\Z_3)$ gapped boundaries.}
Let us now specialize to $\mfD(\Z_3)$, or the bilayer $\nu=1/3$ FQH. Ref. \cite{Bark16} proposed to use superconducting ($1+e+\overbar{e}$) boundaries to encode qutrits, so the read out can be done with electric charge measurement. We follow this scheme, and occasionally use the other ($m$-boundary) encoding as an ancilla.

By Ref. \cite{Cui15-m}, one universal qutrit gate set is the {\it metaplectic gate set}: 
\begin{enumerate}[nolistsep]
\item
The single-qutrit Hadamard gate $H_3$.
\item
The two-qutrit entangling gate $\SUM_3$.
\item
The single-qutrit generalized phase gate
$Q_3 = \text{diag}(1,1,\omega)$.
\item
Any nontrivial single-qutrit classical (i.e. Clifford) gate not equal to $H_3^2$.
\item A projection $M$ of a state in the qutrit space $\C^3$ to $\Span\{ \ket{0} \}$ and its orthogonal complement $\Span\{ \ket{1}, \ket{2} \}$, so that the resulting state is coherent if projected into $\Span\{ \ket{1}, \ket{2} \}$.
\end{enumerate}

\begin{figure}
\centering
\includegraphics[width = 0.4\columnwidth]{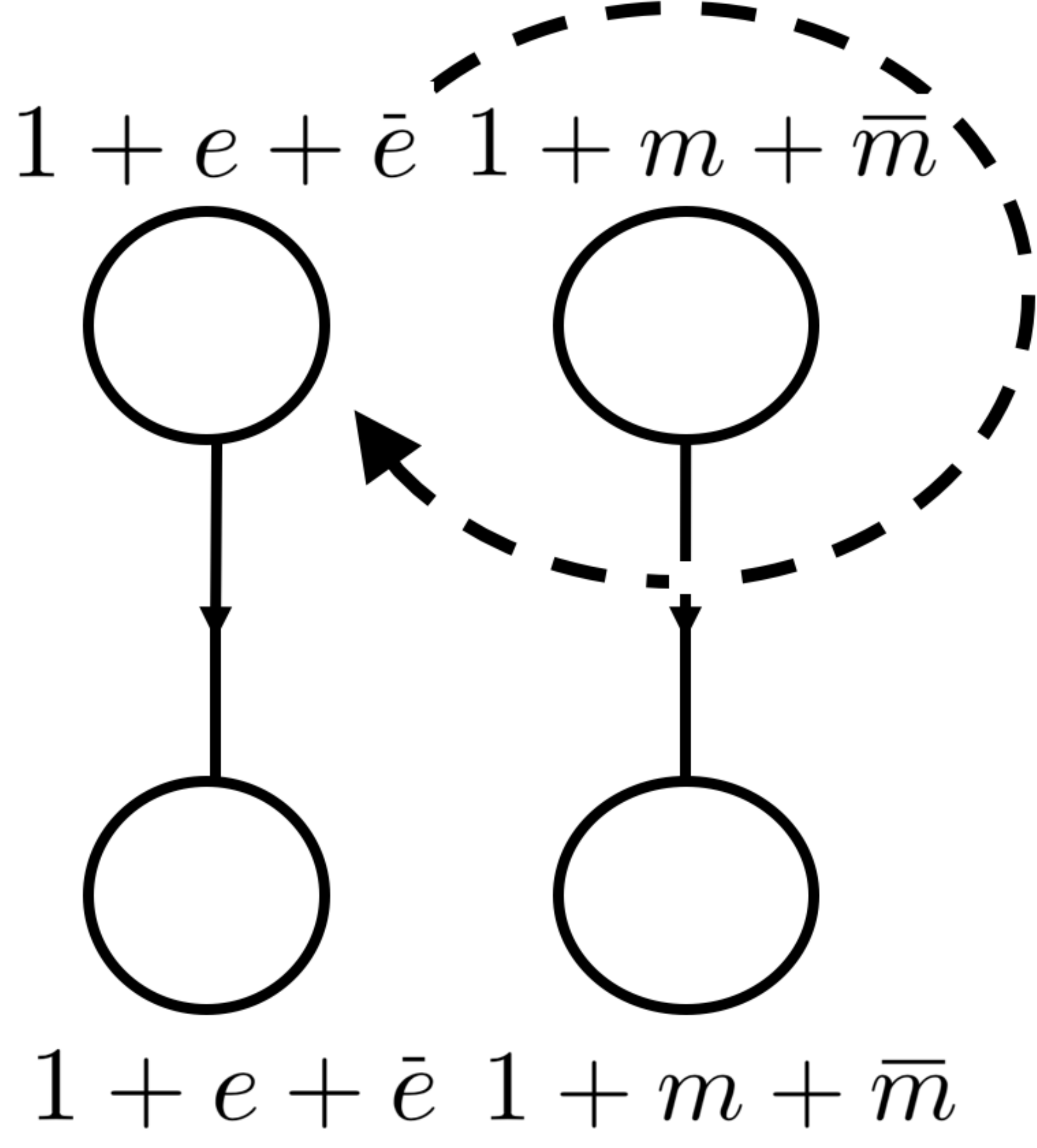}
\caption{Braid for the $\wedge\sigma^z_3$ gate.}
\label{fig:dz3-braiding}
\end{figure}

\begin{figure}
\centering
\includegraphics[width=0.9\textwidth]{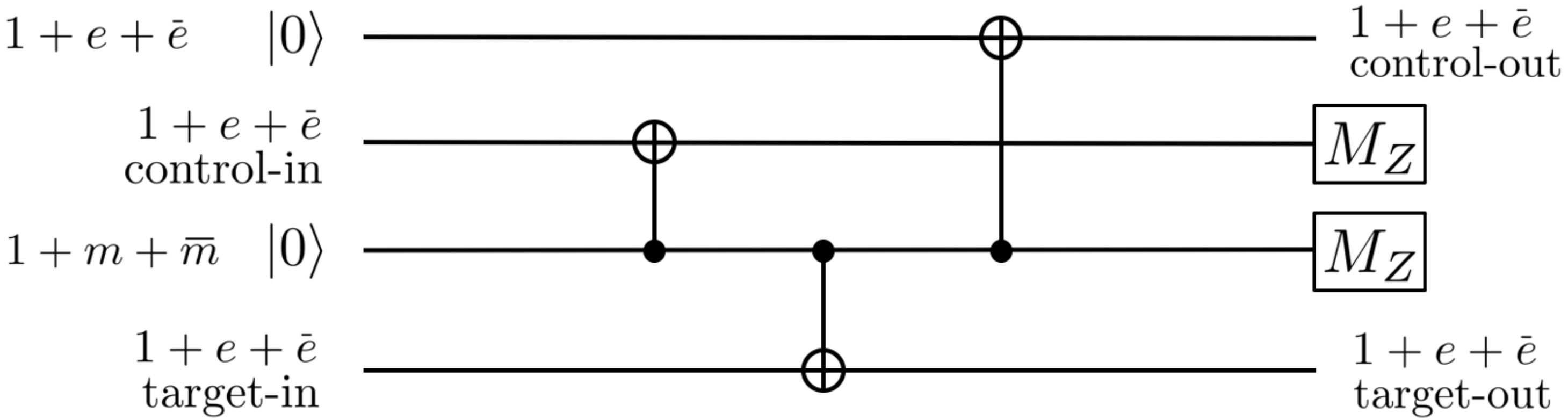}
\caption{Short circuit (generalizing Ref. \cite{Fowler12}) to use three SUM gates between $1+m+\overbar{m}$ and $1+e+\overbar{e}$ qutrits to implement a SUM gate between $1+e+\overbar{e}$ qutrits. All entangling gates drawn are $\SUM_3$.}
\label{fig:dz3-sum}
\end{figure}

We now discuss how each of these gates can be implemented from the aforementioned topological operations. First, we discuss the implementation of 1-4:
\begin{enumerate}[nolistsep]
	\item $H_3$ is equal to the modular $\mathcal{S}$ matrix of the anyon theory $\mathrm{SU}(3)_1$, so it is in the representation of mapping class group of the torus and can be implemented via a sequence of topological charge projections.
	\item For $\SUM_3$, consider braiding one hole of a $e$-boundary target qutrit with another hole of a $m$-boundary control qutrit (i.e. apply $\sigma_2^2$, as shown in Fig. \ref{fig:dz3-braiding}). This gives
\begin{equation}
\sigma_2^2 = \text{diag} (1,1,1,1,\omega,\omega^2,1,\omega^2,\omega) = \wedge \sigma^z_3.
\end{equation}

Because we implemented the Hadamard and $\wedge \sigma^z_3  = (I_3 \otimes H_3) \SUM_3 (I_3 \otimes H_3)$, conjugating the target qutrit by Hadamards gives the SUM gate between a $1+e+\overbar{e}$ qutrit and a $1+m+\overbar{m}$ qutrit. We then have a short circuit (Fig. \ref{fig:dz3-sum}) using these SUM gates to implement a SUM gate between two $1+e+\overbar{e}$ qutrits. After the circuit, one must interpret the measurement outcome of the ancilla qutrit. If we measure $\ket{m^j}$, we must apply $(\sigma^x_3)^j$ to the control-out (e.g. by applying $W_{e^j}(\gamma)$). 
\item By Ref. \cite{Barkeshli16}, topological charge projections can be used to implement $\mathrm{diag}(\mathrm{1,\omega, \omega})$, the Dehn twist of the $\mathrm{SU}(3)_1$ theory. We follow this by a generalized Pauli-Z gate to obtain $Q_3$.
\item By Eq. (\ref{eq:tunnel-formula}), the tunneling operator $W_e (\gamma)$ implements the single-qutrit Pauli-X gate $\sigma^x_3$. 
\end{enumerate}

The implementation of the coherent projection $M$ is the most challenging part of the proposal. First, we relate $M$ to a TCM. A planar $\mfD(\Z_3)$ with two $1+e+\overbar{e}$ boundaries can be viewed as double layers of $\mathrm{SU}(3)_1$ connected via two handles; the curve $\gamma$ connecting the two boundaries lifts to a loop in this perspective. By Eq. (\ref{eq:tcp}), projecting to vacuum within this loop gives
\begin{equation}
P_\gamma^{(1)} = \frac{1}{3}
\begin{bmatrix}
1 & 1 & 1 \\
1 & 1 & 1 \\
1 & 1 & 1
\end{bmatrix}.
\end{equation}
\noindent
The eigenvalues and eigenspaces of $P_\gamma^{(1)}$ are:
\begin{equation}
\lambda = 0: \Span \left\{ 
\begin{bmatrix}
1 \\
\omega \\
\overbar{\omega}
\end{bmatrix},
\begin{bmatrix}
1 \\
\overbar{\omega} \\
\omega
\end{bmatrix}
\right\}
\quad
\lambda = 1: \Span \left\{ 
\begin{bmatrix}
1 \\
1 \\
1
\end{bmatrix}
\right\}
\end{equation}

One then obtains the coherent projection $M$ by conjugating the orthogonal projector $1-P_\gamma^{(1)}$ with the Hadamard, i.e. $H^{\dagger}_3 (1-P_\gamma^{(1)})H_3$.  While $P_\gamma^{(1)}$ is a topological charge projection as in Ref. \cite{Barkeshli16}, $1-P_\gamma^{(1)}$ is a general TCM. 



We now have universal quantum computation using gapped boundaries of $\mfD(\Z_3)$. This is very significant, as we achieve universal quantum computation using only an abelian TQFT (all anyon braidings in $\mfD(\Z_3)$ are projectively trivial), without using state injection, as in Ref. \cite{Fowler12}. 

\vspace{1mm}
\noindent
{\it Symmetry-protected realization.} In physical realizations such as bilayer FQH, the TCM can be implemented as follows: we tune the system such that the quasiparticle tunneling along the desired loop is enhanced, so that the system has the projected charge state as the ground state. This can be achieved by e.g. using gate configurations to diminish the energy gap. We consider $1-P_\gamma^{(a)}$ as a concrete example. The desired term in the Hamiltonian we would like to create is $H'=-t W_\gamma(e)+\text{h.c.}$ where $t$ is the (complex) tunneling amplitude and $W_\gamma(e)$ is the Wilson tunneling operator. $W_\gamma$ has eigenvalues $1, \omega, \bar{\omega}$. The coherent projection requires that the eigenvalues of $H'$ split into two sets, one of which has two \emph{degenerate} eigenvalues. This puts a stringent constraint on the complex phase of $t$. The simplest choice is that $t$ is real.

The requirement that $t$ is real is beyond topological protection. Physically, such condition can be met in a  fractional quantum spin Hall state~\cite{Levin09, Levin12}, an interacting analog of quantum spin Hall insulator enriched by time-reversal symmetry. Topologically, this phase is identical to bilayer $\nu=1/3$ Laughlin state, if the layer index is actually identified as the electron spin up and down. In such a state, the time-reversal symmetry exchanges the two layers. The $e$ anyon in this physical realization is the bound state of the spin up/down quasiholes. Therefore, the tunneling amplitude of $e$ has to be real since $e$ is time-reversal invariant, and the TCM is symmetry-protected.

\noindent
{\bf Conclusions.} Gapped boundaries provide the missing $\frac{\pi}{8}$-gate for a universal gate set from the doubled Ising theory \cite{Barkeshli16}.  In this Letter, we use our symmetry-protected TCM to obtain a coherent projection, which augments the topological operations from Ref. \cite{Bark16} for the $\Z_3$ toric code to a universal gate set for a qutrit computational model. The $\Z_3$ toric code is realized by bilayer fractional quantum Hall liquids \cite{Bark16},  whereas it is not yet clear how to physically realize the doubled Ising theory.  The challenge for a realistic implementation of our universal  gate set now lies in a practical realization of the coherent projection.

\begin{acknowledgments}
{\bf Acknowledgment.} The authors thank Maissam Barkeshli, Shawn Cui, and C{\'e}sar Galindo for answering many questions. I.C. would like to thank Michael Freedman and Microsoft Station Q for hospitality in hosting the summer internship and visits during which this work was done.  Z.W. is partially supported by NSF grant DMS-1411212.
\end{acknowledgments}


\end{document}


\begin{appendix}

\vspace{2mm}

\section{Appendix A: Notations}
\label{sec:notations}

In this Appendix, we list all of the notations that are used throughout the paper.

The Drinfeld center of a category is denoted $\mZ(\mC)$. When $\mC$ is the representation category of a finite group $G$, we have $\mfD(G) = \Rep(D(G)) = \mZ(\Rep(G)) = \mZ(\text{Vec}_G)$.

We will adopt the following conventions for labeling anyons, gapped boundaries, and their excitations:

\begin{enumerate}
\item
Bulk excitations (a.k.a. anyons or topological charges), which are the simple objects within the modular tensor category $\B = \ZZ(\text{Rep}(G))$ or $\B = \ZZ(\CC)$ will be labeled by $a,b,c...$. Their dual excitations are labeled by $\overbar{a}, \overbar{b}, \overbar{c}, ...$, respectively.
\item
The gapped boundary will be given as a Lagrangian algebra $\A$ which is an object in $\B$.
\item
Excitations on the boundary will be labeled as $\alpha, \beta, \gamma, ...$. When necessary, the local degrees of freedom during condensation will be labeled as $\mu, \nu, \lambda, ...$.
\end{enumerate}

Furthermore, when using any $F$ symbols and $R$ symbols for a fusion category or a modular tensor category, we will adopt the following conventions for indices:

\begin{equation}
\vcenter{\hbox{\includegraphics[width = 0.4\textwidth]{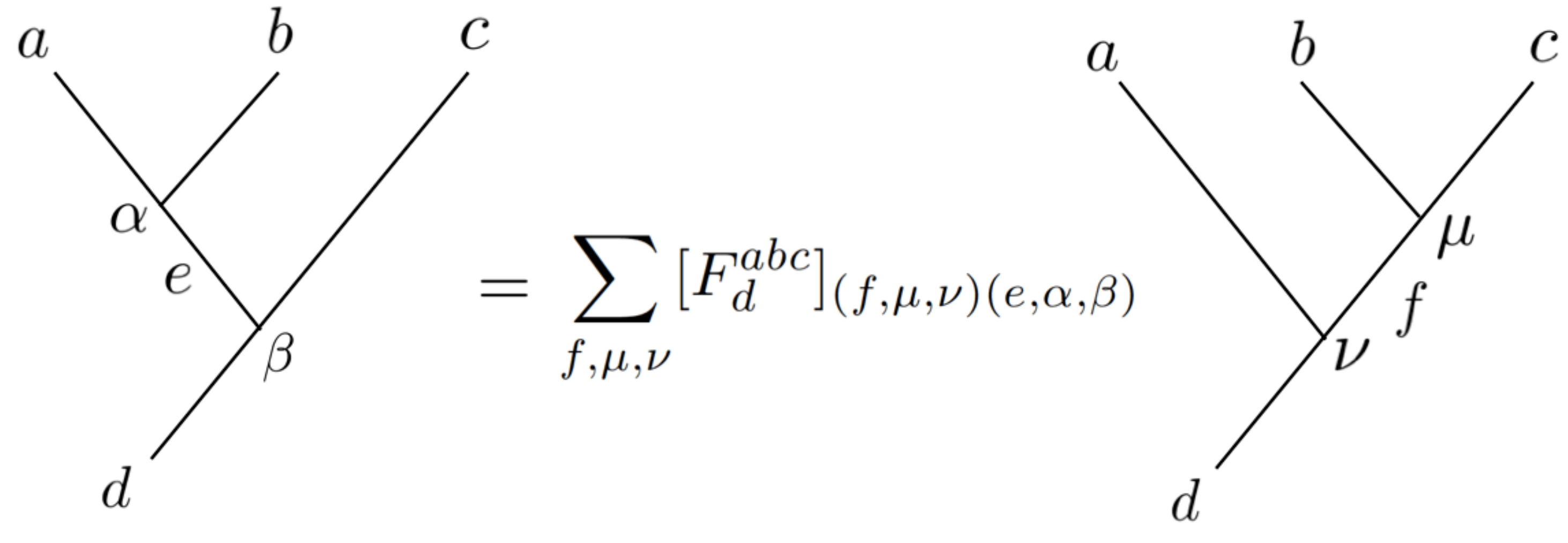}}}
\end{equation}

\begin{equation}
\vcenter{\hbox{\includegraphics[width = 0.26\textwidth]{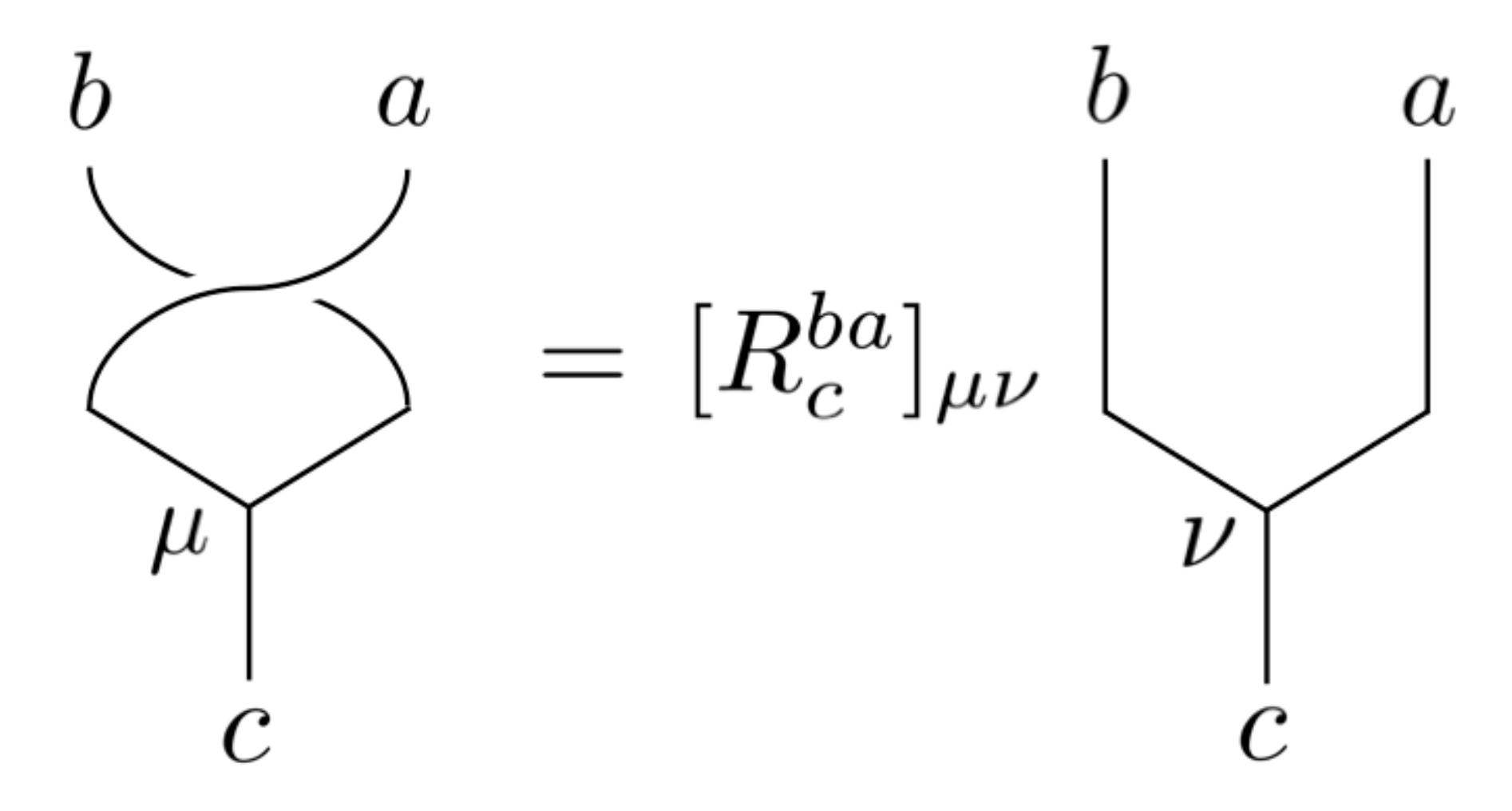}}}.
\end{equation}

Given a gapped boundary $\A$ and anyons $a,b,c$ that condense to vacuum on the boundary, the associated $M-3j$ symbols are defined as in Refs. \cite{Cong16a,Cong16b}:

\begin{equation}
\vcenter{\hbox{\includegraphics[width = 0.3\textwidth]{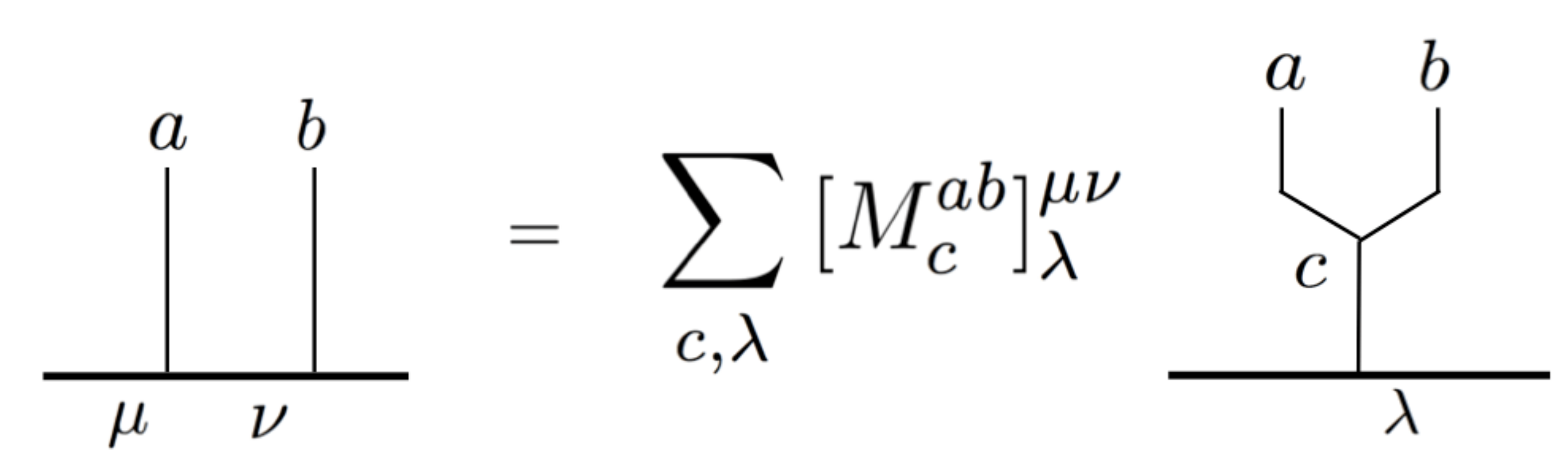}}}
\end{equation}

\noindent
These $M$ symbols encode the associativity of bulk anyon fusion and condensation to vacuum on the boundary.

\vspace{2mm}
\section{Appendix B: Deriving formulas for the topologically protected operations}
\label{sec:derivations}

\subsection{Tunnel-$a$ operations}

Let us consider an arbitrary basis element $W_b(\gamma)\ket{0}$ of the qudit. Diagrammatically, after applying the $W_a(\gamma)$ operator, we have arrived in the following state:

\begin{equation}
\label{eq:tunnel-1}
\vcenter{\hbox{\includegraphics[width = 0.25\textwidth]{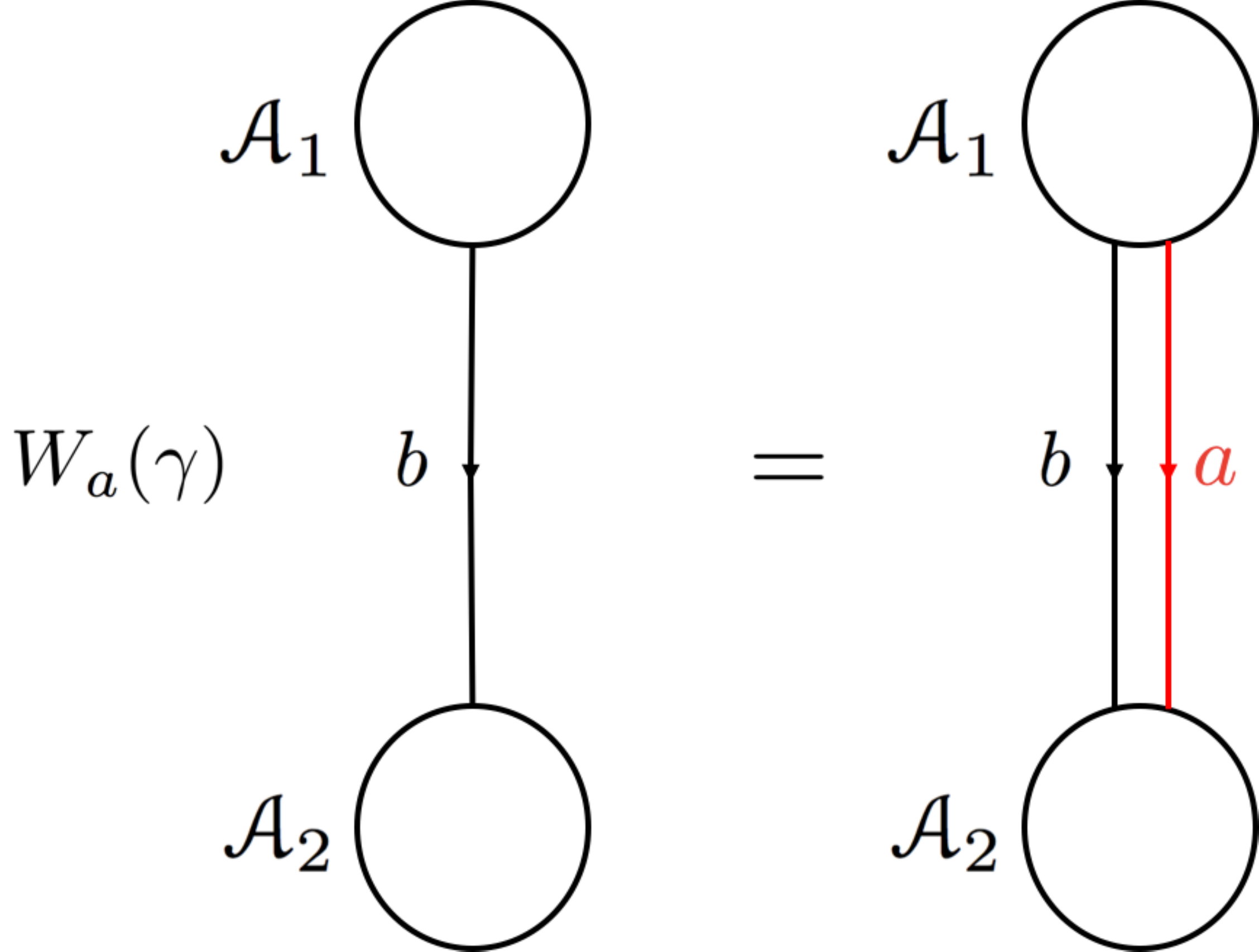}}}
\end{equation}

\noindent
Here, and for the rest of the section, solid black lines are used to indicate a basis element of the hom-space that describes the ground state, while solid red lines are used to denote Wilson operators.

To express this in terms of our original basis, we must convert the two anyon-tunneling ribbon operators into one. To do this, we can first apply the $M$-3$j$ operator and its Hermitian conjugate to the bottom and top boundaries of (\ref{eq:tunnel-1}), respectively, to get\footnote{In this analysis, we will drop the multiplicity indices $\mu,\nu,\lambda$ from the $M$ symbols for concision; the generalization is obvious.}:

\begin{equation}
\label{eq:tunnel-2}
\vcenter{\hbox{\includegraphics[width = 0.26\textwidth]{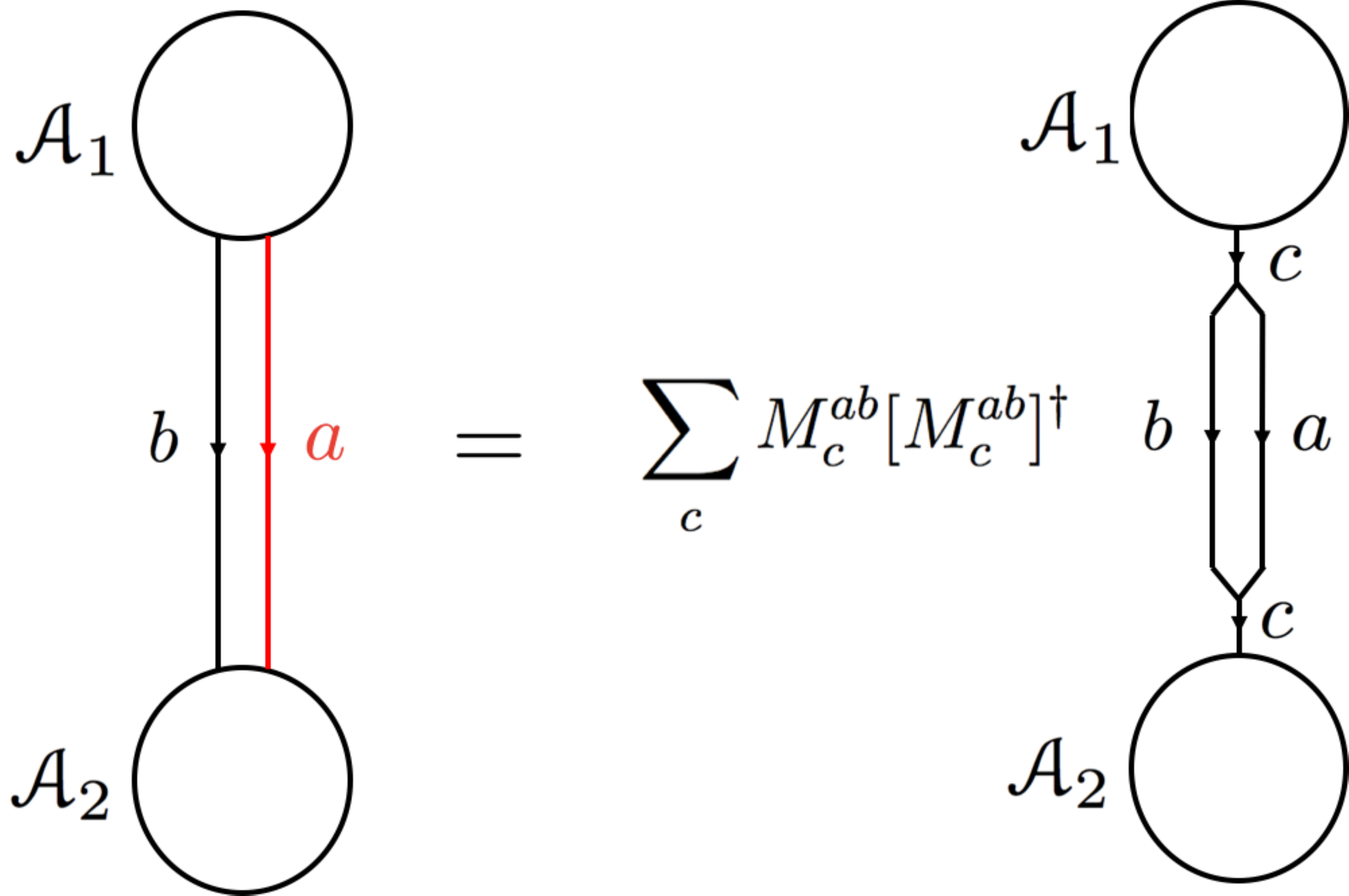}}}
\end{equation}

\noindent
Here, $M^{ab}_c(\A_i)$ indicates that the $M$-3$j$ symbol is for the gapped boundary given by the Lagrangian algebra $\A_i$.

We are now left with a bubble in the bulk. This can be eliminated using $\theta$ symbols of the bulk modular tensor category, by the following relation:

\begin{equation}
\label{eq:tunnel-3}
\includegraphics[width = 0.26\textwidth]{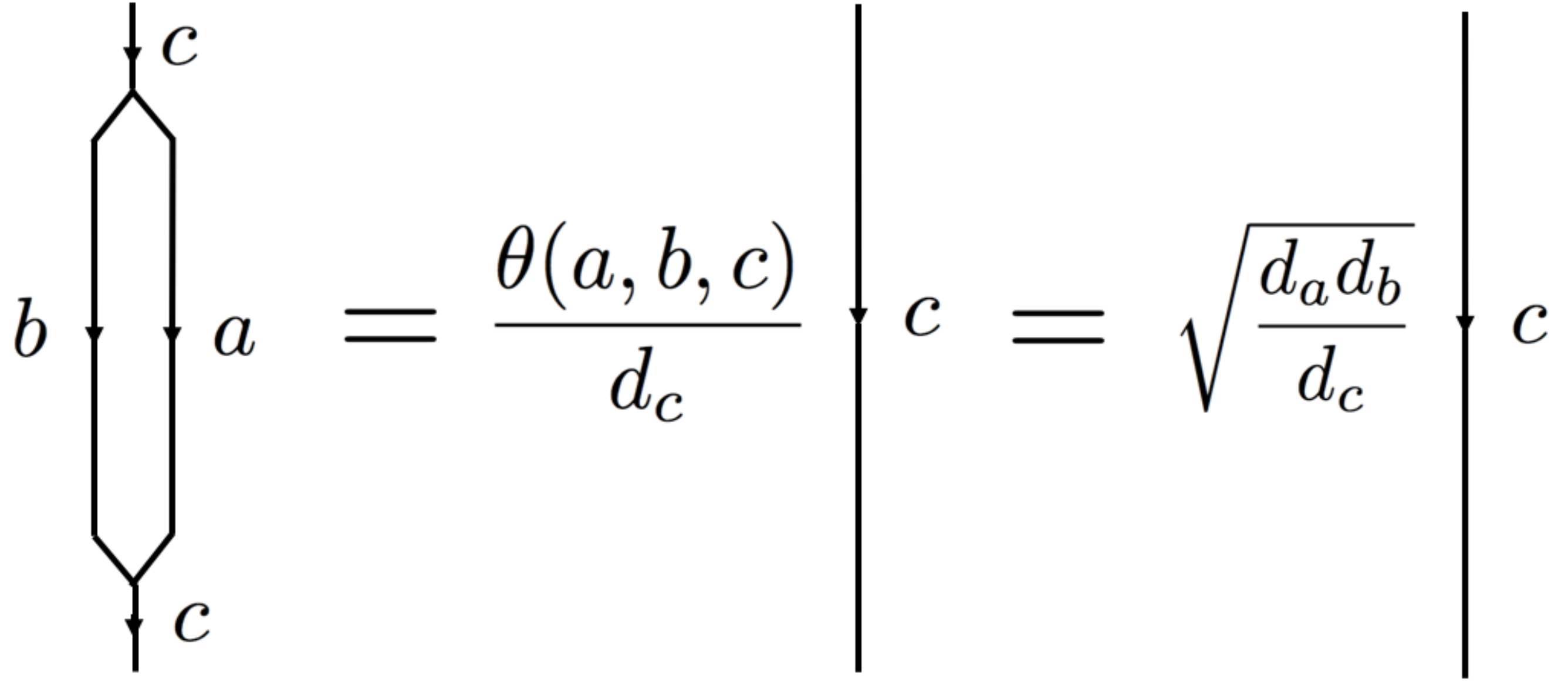}
\end{equation}

Hence, we have the following equation:

\begin{multline}
W_a(\gamma) W_b(\gamma) \ket{0} =
W_a(\gamma) \vcenter{\hbox{\includegraphics[width = 0.06\textwidth]{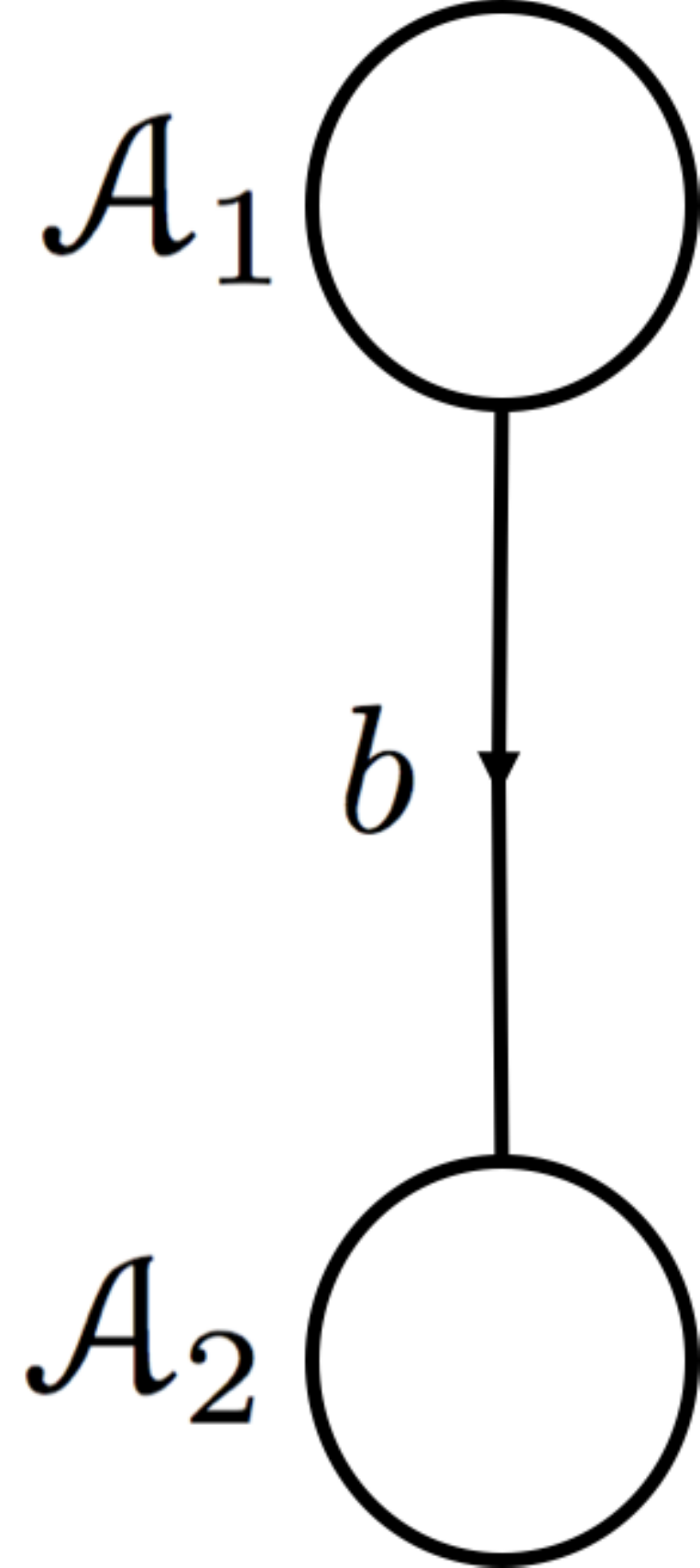}}} \\
= \sum_c M^{ab}_c (\A_1) [M^{ab}_c]^\dagger (\A_2) \sqrt{\frac{d_a d_b}{d_c}}
\vcenter{\hbox{\includegraphics[width = 0.06\textwidth]{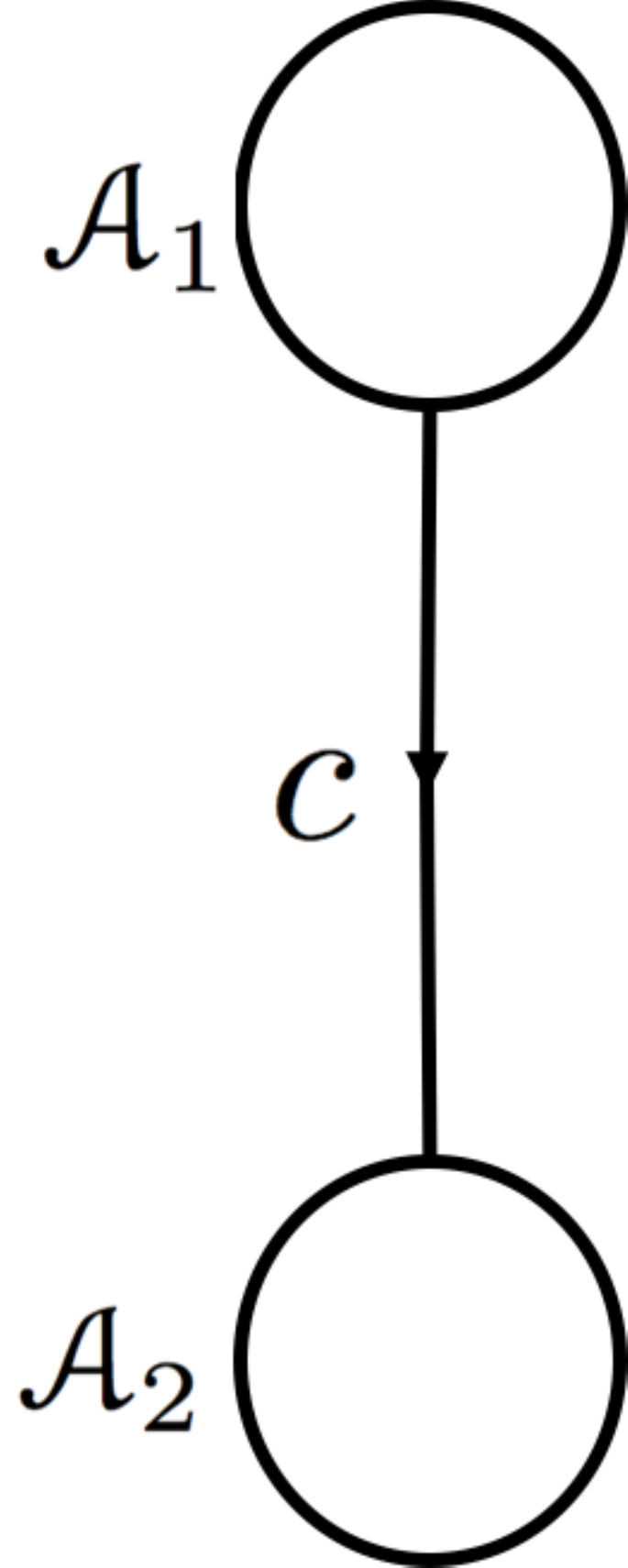}}} 
\end{multline}

In general, $W_a(\gamma)$ is unitary if and only if $a$ cannot condense to any (non-vacuum) excitation on the boundary.

\subsection{Loop-$a$ operations}
\label{sec:loop}

Another topological operation we can consider is to create a pair of anyons $a,\overbar{a}$ in the bulk, move $a$ counter-clockwise around a gapped boundary, and come back and annihilate the pair to vacuum. Physically, this corresponds to applying the $a$ ribbon operator to a counter-clockwise closed ribbon encircling the gapped boundary. We will denote this operation by $W_a(\alpha_i)$, where $\alpha_i$ is the closed ribbon encircling boundary $i$. This operator is often known as the {\it Wilson loop operator}.

Suppose we have two gapped boundaries given by Lagrangian algebras $\A_1,\A_2$, which encode a qudit with orthonormal basis as in Fig. \ref{fig:algebraic-gsd-n-2}. As before, we would like to compute the result of applying each $W_a(\alpha_i)$ on each basis element of the ground state $\Hom(1, \A_1 \otimes \A_2)$, and express the result in terms of the original basis.

Suppose we start as an arbitrary basis element $W_b(\gamma)\ket{0}$. Diagrammatically, the operator $W_a(\alpha_2)$ transforms this basis element into the following state:

\begin{equation}
\label{eq:loop-1}
\vcenter{\hbox{\includegraphics[width = 0.24\textwidth]{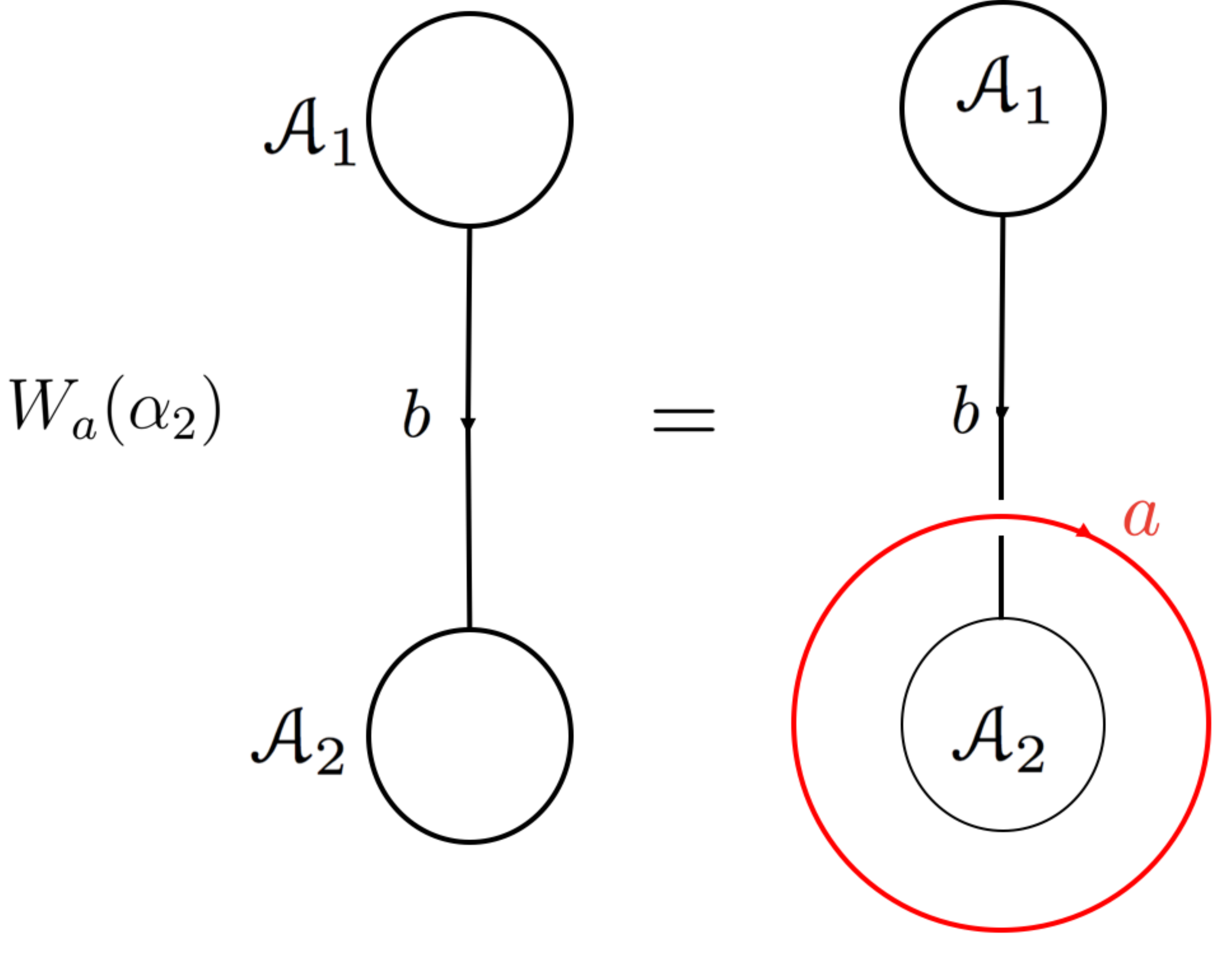}}}
\end{equation}

\noindent
Since we are working in a model where the total charge is vacuum, we may consider this picture as two holes on a sphere. We can hence push the anyon loop back through infinity, to get

\begin{equation}
\label{eq:loop-2}
\vcenter{\hbox{\includegraphics[width = 0.24\textwidth]{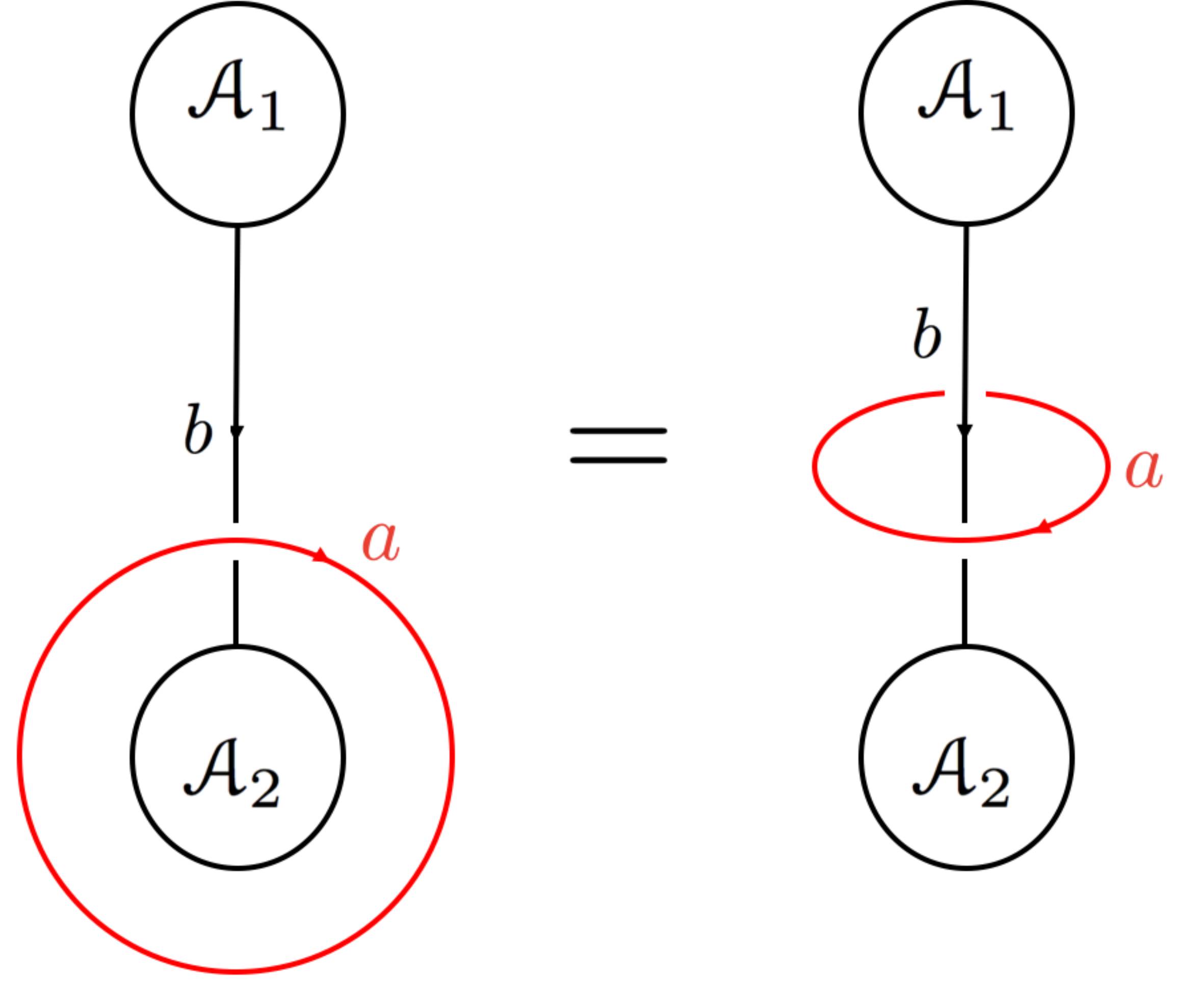}}}
\end{equation}

\noindent
The right hand side of Eq. (\ref{eq:loop-2}) may be simplified using the definition of the $\mathcal{S}$ matrix as follows: Suppose

\begin{equation}
\label{eq:loop-3}
\vcenter{\hbox{\includegraphics[width = 0.2\textwidth]{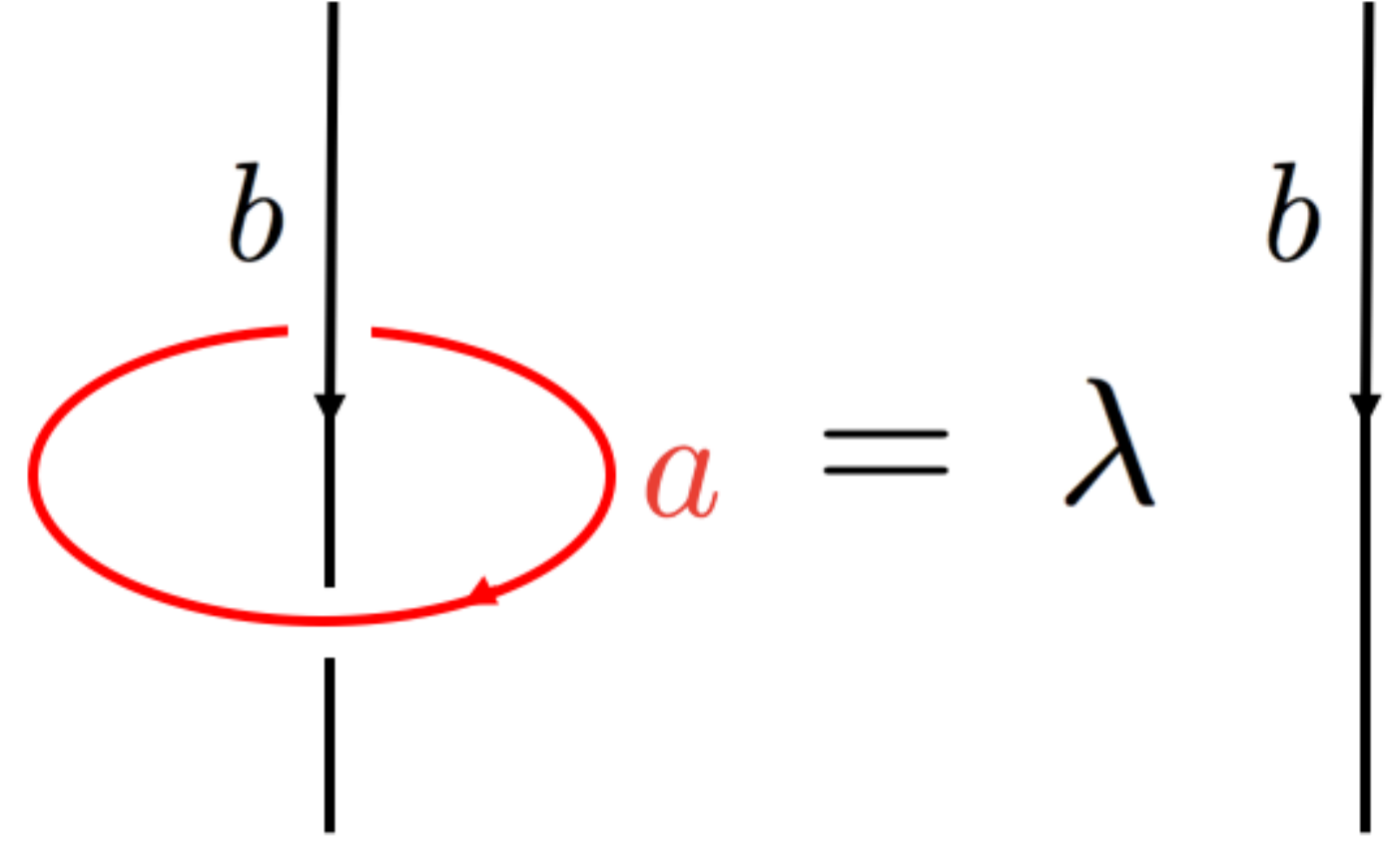}}}.
\end{equation}

\noindent
Then, taking traces on both sides, we get

\begin{equation}
\label{eq:loop-4}
\vcenter{\hbox{\includegraphics[width = 0.38\textwidth]{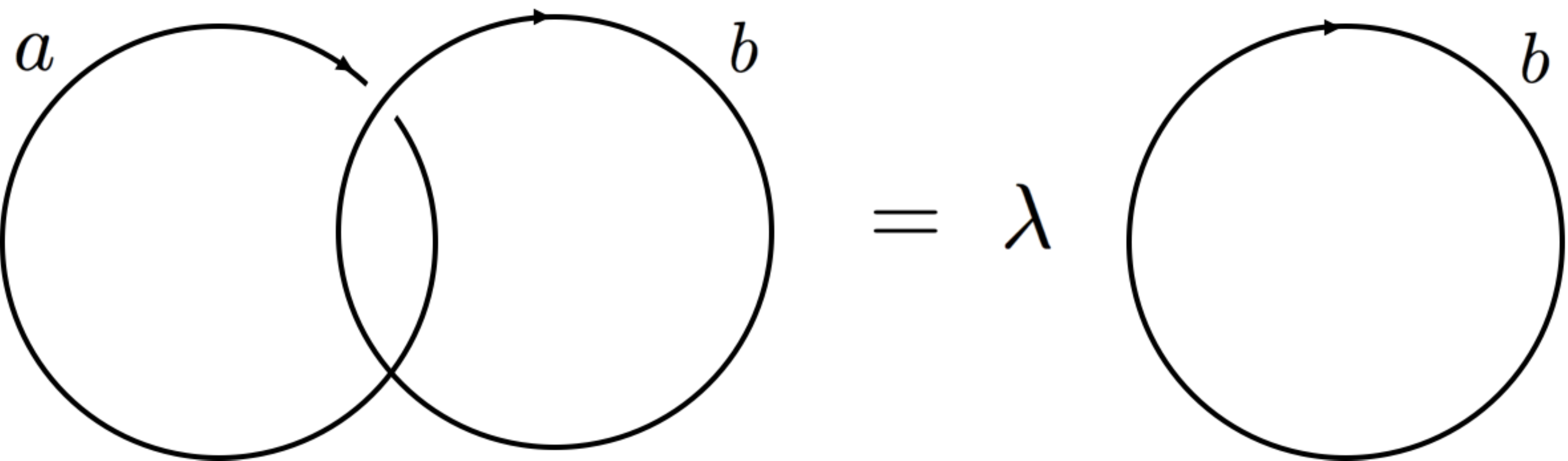}}}.
\end{equation}

By definition of the $\mathcal{S}$ matrix of the modular tensor category $\B$ \cite{BakalovKirillov}, we have $\lambda = \frac{S_{ab}}{d_b}$. Hence, we have the formula

\begin{equation}
W_a(\alpha_2) \vcenter{\hbox{\includegraphics[width = 0.06\textwidth]{ch5_basis}}} = \frac{S_{ab}}{d_b}  \vcenter{\hbox{\includegraphics[width = 0.06\textwidth]{ch5_basis}}}
\end{equation}

As before, $W_a(\alpha_i)$ gives a $d\times d$ matrix that acts on the ground state $\Hom(1, \A_1 \otimes \A_2)$.

As in the case of the tunneling operator, the loop operator $W_{a}(\alpha_i)$ also need not be Hermitian or unitary. In general, it is just a Wilson loop operator, which is a holonomy.

\subsection{Braiding gapped boundaries}

We now present the derivation for the formula of $\sigma_2^2$. 
Consider an arbitrary basis element of $\Hom(\one_\B, \A_1 \otimes \A_2 \otimes \A_3 \otimes \A_4)$ as our start state. After applying $\sigma_2^2$, we have\footnote{For the rest of this derivation, we assume for simplicity of illustration and computation that all anyons are self-dual. The generalization is obvious, but one just needs to be more careful in drawing orientations for each edge and using $F$ and $R$ symbols.}:

\begin{equation}
\label{eq:braid-1}
\sigma_2^2 \vcenter{\hbox{\includegraphics[width = 0.11\textwidth]{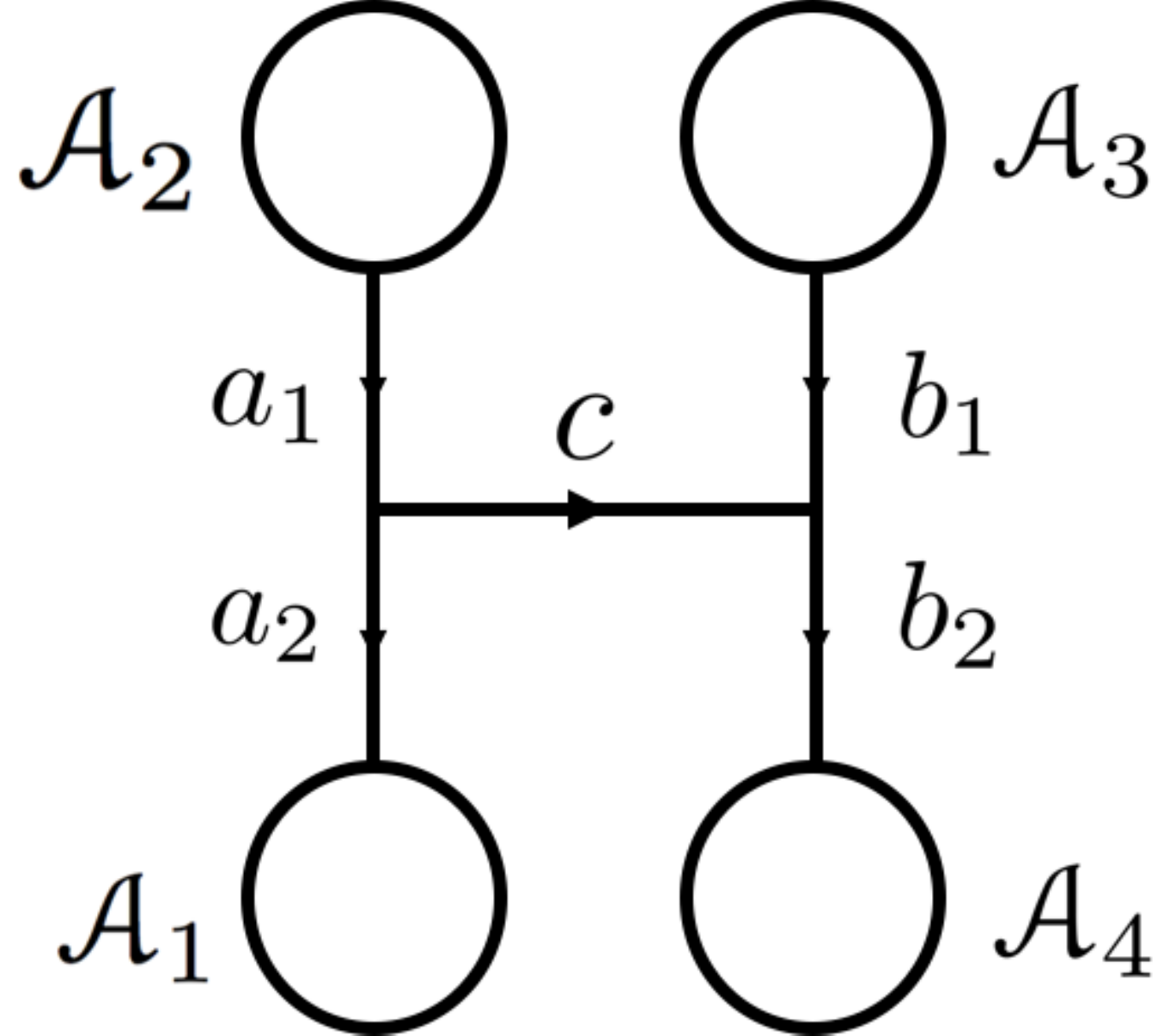}}} = \vcenter{\hbox{\includegraphics[width = 0.12\textwidth]{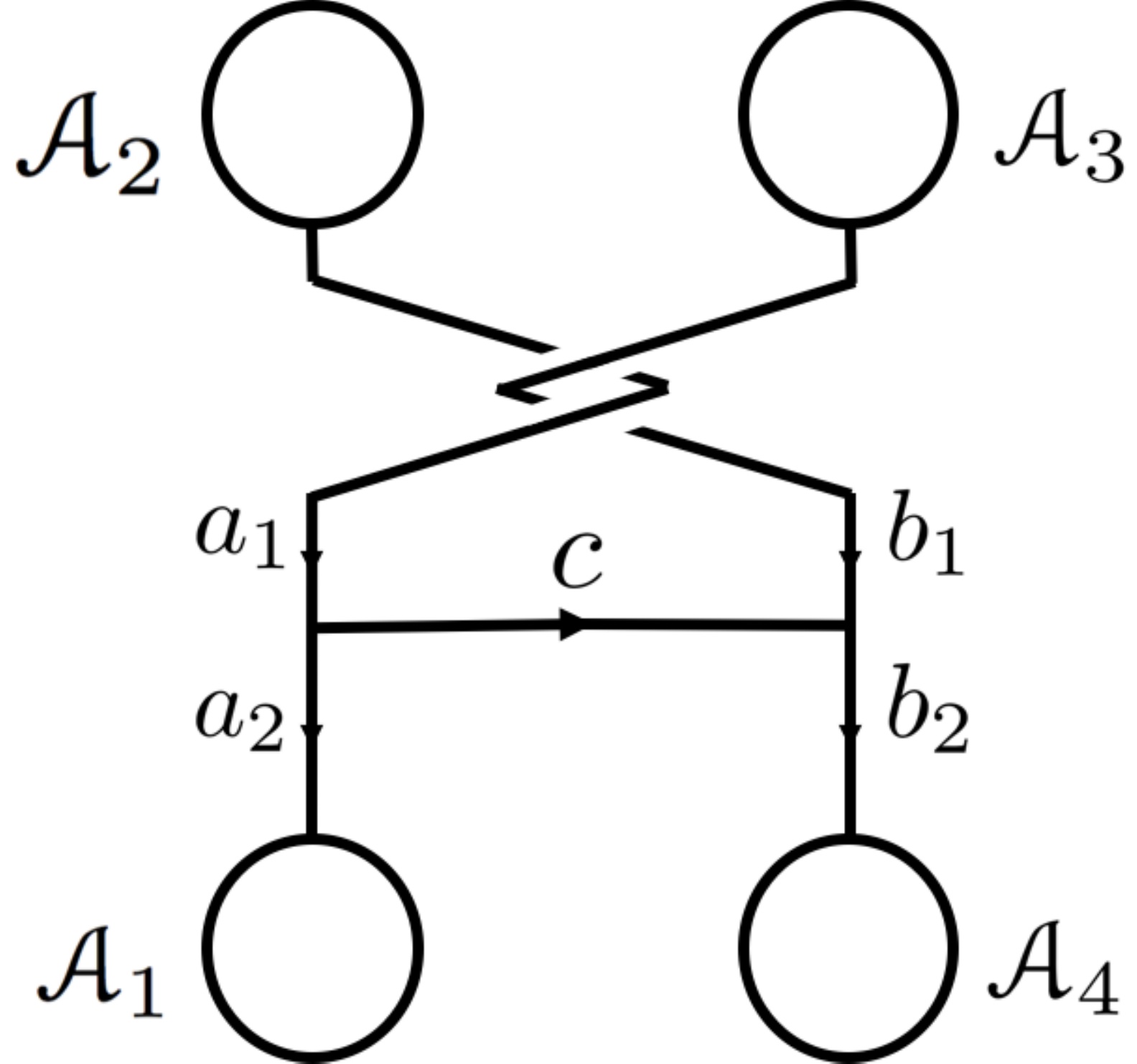}}}
\end{equation}

\noindent
As with the earlier cases, we would like to express the right hand side of Eq. (\ref{eq:braid-1}) in terms of the original basis. We can apply an $F$-move to get:

\begin{equation}
\label{eq:braid-2}
\vcenter{\hbox{\includegraphics[width = 0.12\textwidth]{ch5_3_braid_2}}} = \sum_{c'} F^{a_2 a_1 b_1}_{b_2;c'c}
\vcenter{\hbox{\includegraphics[width = 0.12\textwidth]{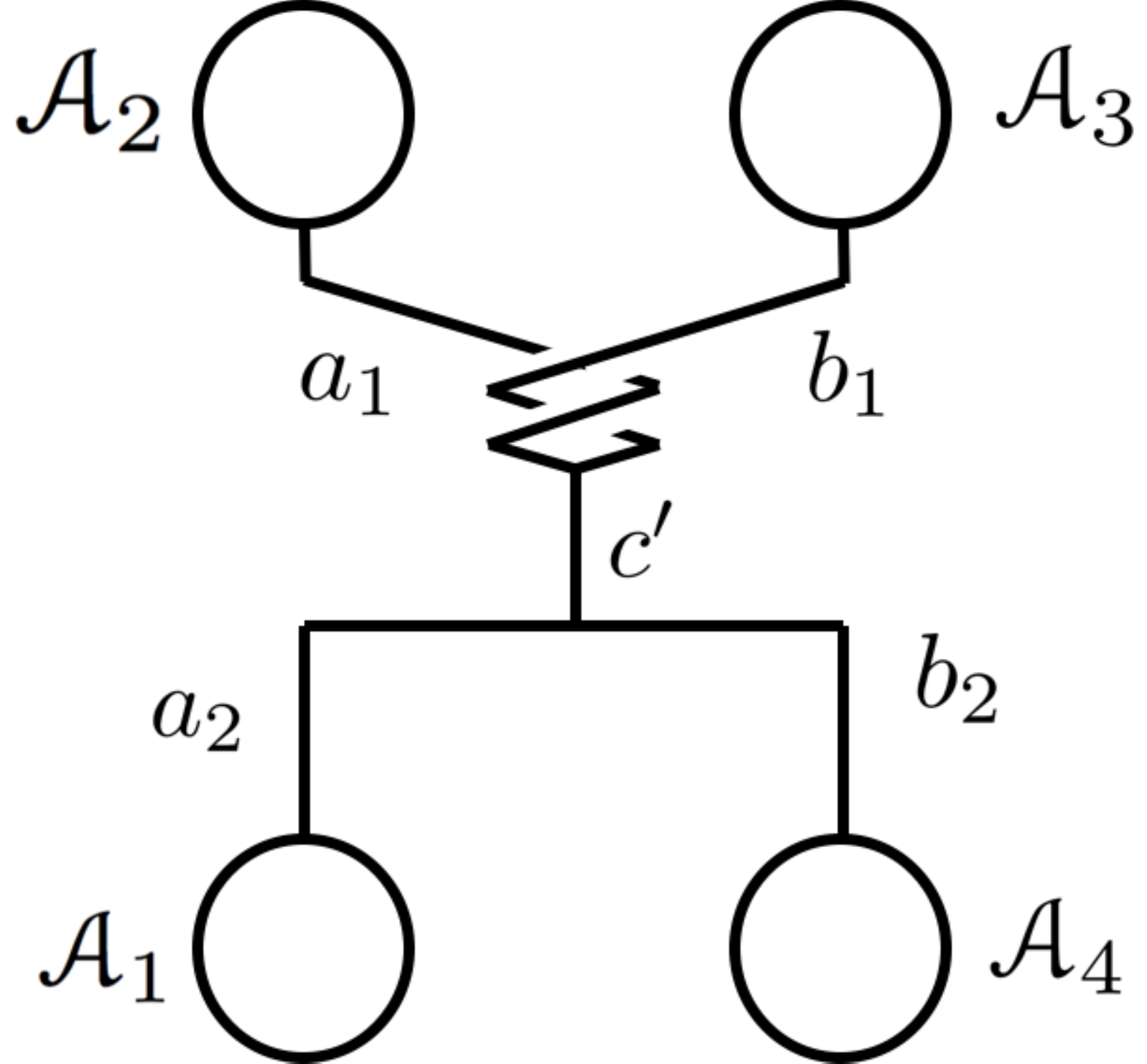}}}
\end{equation}

\noindent
Next, applying two $R$-moves gives:

\begin{equation}
\label{eq:braid-3}
\vcenter{\hbox{\includegraphics[width = 0.12\textwidth]{ch5_3_braid_3}}} = R^{a_1 b_1}_{c'} R^{b_1 a_1}_{c'}
\vcenter{\hbox{\includegraphics[width = 0.12\textwidth]{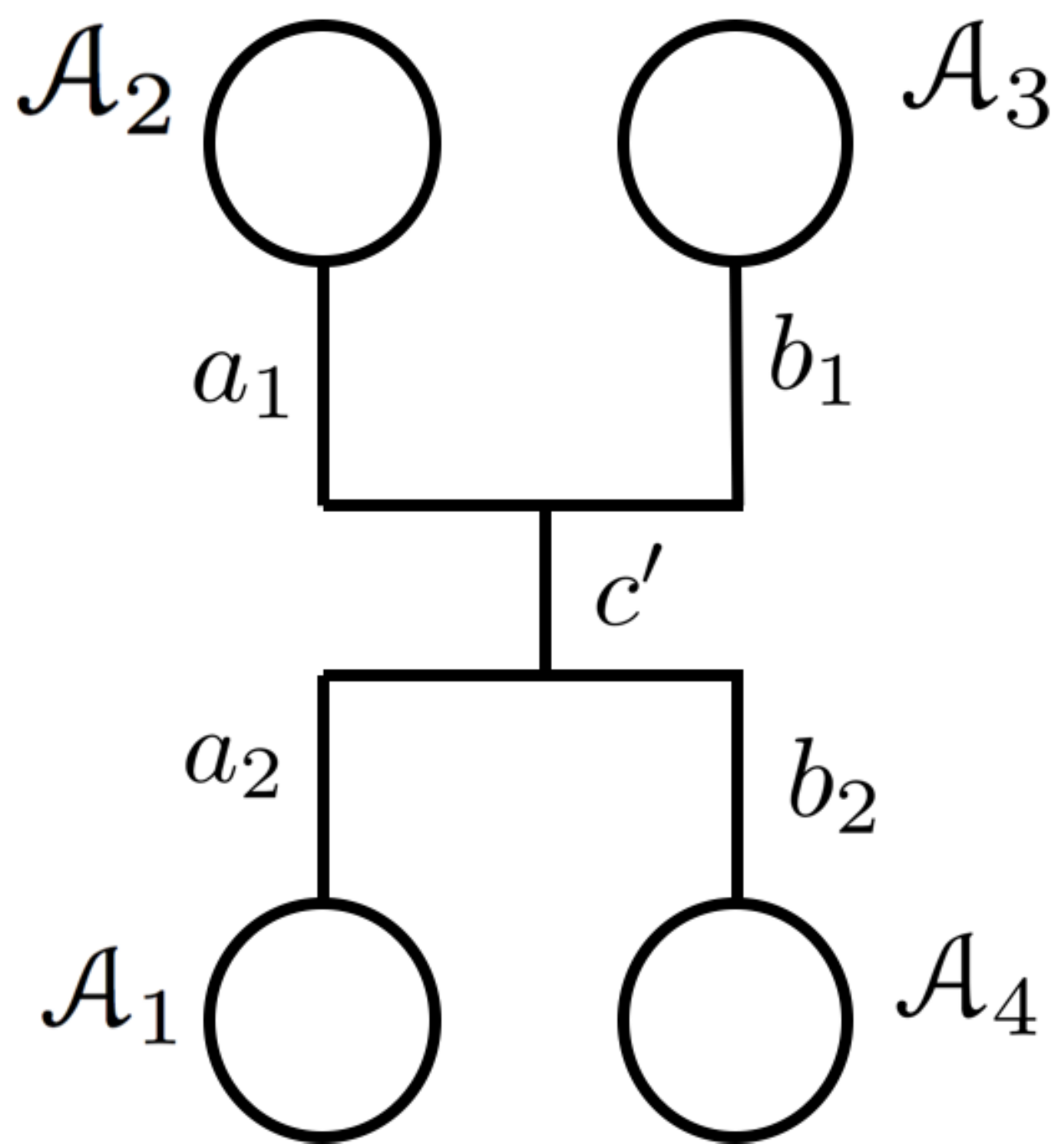}}}
\end{equation}

\noindent
Finally, we apply one more $F$-move, which gives:

\begin{equation}
\label{eq:braid-4}
\vcenter{\hbox{\includegraphics[width = 0.1\textwidth]{ch5_3_braid_4}}} = \sum_{c''} (F^{a_2 a_1 b_1}_{b_2})^{-1}_{c'' c'}
\vcenter{\hbox{\includegraphics[width = 0.1\textwidth]{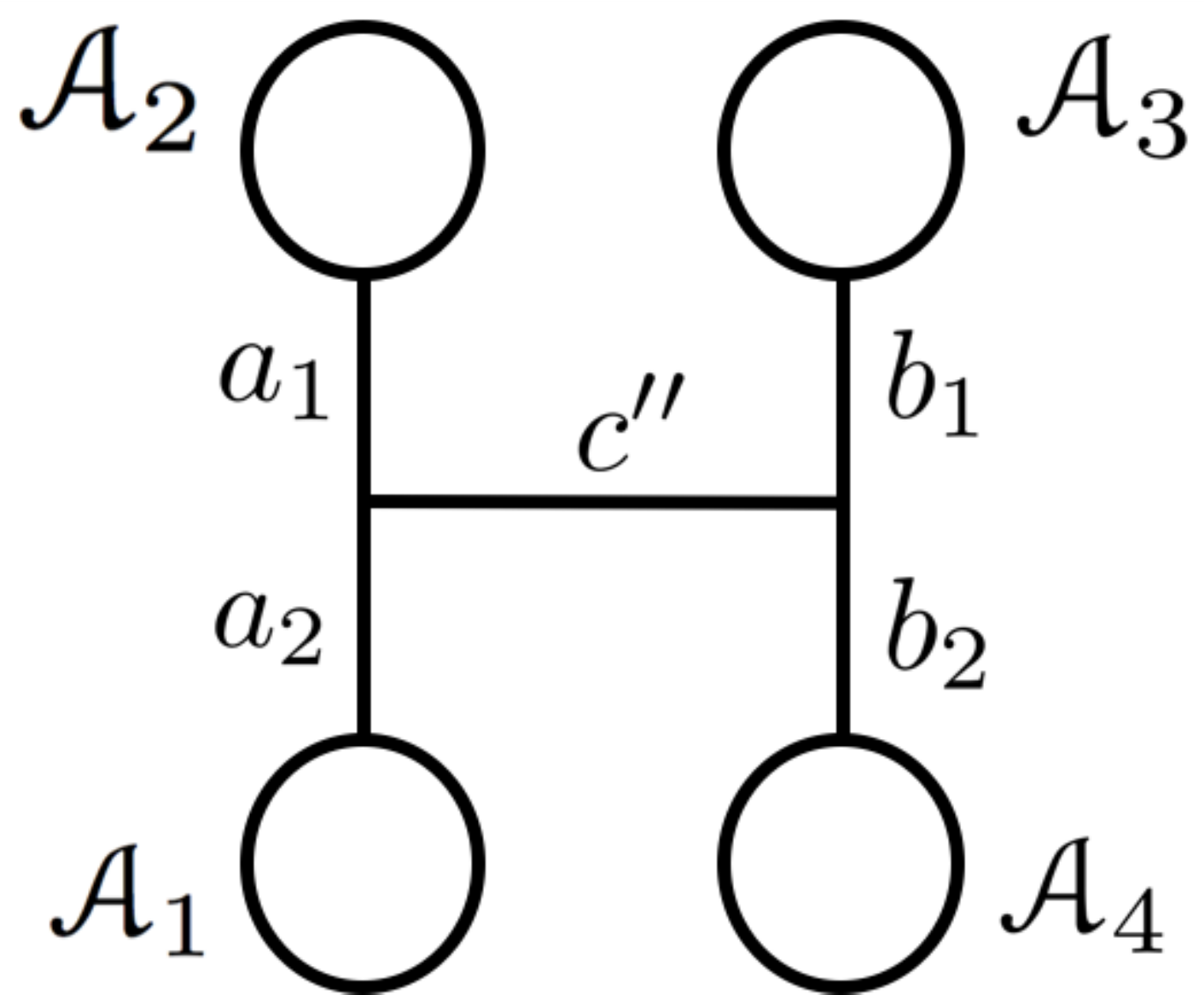}}}
\end{equation}

\noindent
Hence, we see that action of the pure braid group generator $\sigma_2^2$ on an arbitrary basis vector of $\Hom(\one_\B, \A_1 \otimes \A_2 \otimes \A_3 \otimes \A_4)$ is given by the following formula:

{\footnotesize
\begin{equation}
\sigma_2^2 \vcenter{\hbox{\includegraphics[width = 0.09\textwidth]{ch5_3_braid_1}}} = 
\sum_{c,c'} F^{a_2 a_1 b_1}_{b_2;c'c} R^{b_1 a_1}_{c'} R^{a_1 b_1}_c (F^{a_2 a_1 b_1}_{b_2})^{-1}_{c''c'} \vcenter{\hbox{\includegraphics[width = 0.09\textwidth]{ch5_3_braid_5}}}
\end{equation}}

\vspace{2mm}
\section{Appendix C: General definition of topological charge measurement}

If $\B$ is a DW theory, gapped boundary braiding generates only a finite group \cite{Escobar17}. Inspired by the results of Ref. \cite{Barkeshli16}, we introduce topological charge measurement (TCM) based on the {\it Wilson operator algebra} $\mathcal{W}(\B,\{\A_i\})$ for the symmetries of the theory at low energy. 

First, we construct a set $\Gamma(Y)$ of simple loops and arcs in $Y$:

\begin{enumerate}[nolistsep]
\item
For each $i = 1,2, ... n$, let $\alpha_i$ be a simple loop encircling hole $i$ (oriented counter-clockwise). Then $\alpha_i \in \Gamma(Y)$.
\item
	For each pair $1\leq i<j<n$, let $\gamma_{ij}$ be a simple arc connecting hole $i$ and hole $j$ (oriented to point from $i$ to $j$). Then $\gamma_{ij} \in \Gamma(Y)$.
\end{enumerate}

Examples of loops and arcs in $\Gamma(Y)$ are shown in Fig. \ref{fig:tcm}, when $n=2$. Each knot diagram in $Y$ can be resolved using graphical calculus to a linear combination of loops and arcs in $\Gamma(Y)$.

Next, we construct a basis $L_\mW (\B,\{\A_i\})$ for the Wilson operator algebra $\mathcal{W}(\B,\{\A_i\})$:

\begin{enumerate}[nolistsep]
\item
For each anyon $a \in \B$, and for each loop $\alpha_i \in \Gamma(Y)$, the Wilson loop operator $W_{a}(\alpha_i)$ is a basis element.
\item
	For each pair $1\leq i<j<n$, let $A_{i,j}$ be the set of all anyon types $a \in \B$ such that $a$ (resp. $\overbar{a}$) condenses on the $j$-th (resp. $i$-th) boundary. Then, for each $a \in A_{i,j}$, the Wilson line operator $W_{a}(\gamma_{ij})$ is a basis element.
\end{enumerate}

We posit that any Hermitian operator $\mathcal{O}\in \mathcal{W}(\B,\{\A_i\}) = \Span (L_\mW (\B,\{\A_i\}))$ can be measured. Such operators $\mathcal{O}$ are called {\it topological charge measurement operators}.  The corresponding projective measurements $P_\mathcal{O}$ are called {\it topological charge measurements} (TCM).  To be physical, we consider only operators which are monomials of basis operators. 

\begin{figure}
\centering
\includegraphics[width = 0.52\textwidth]{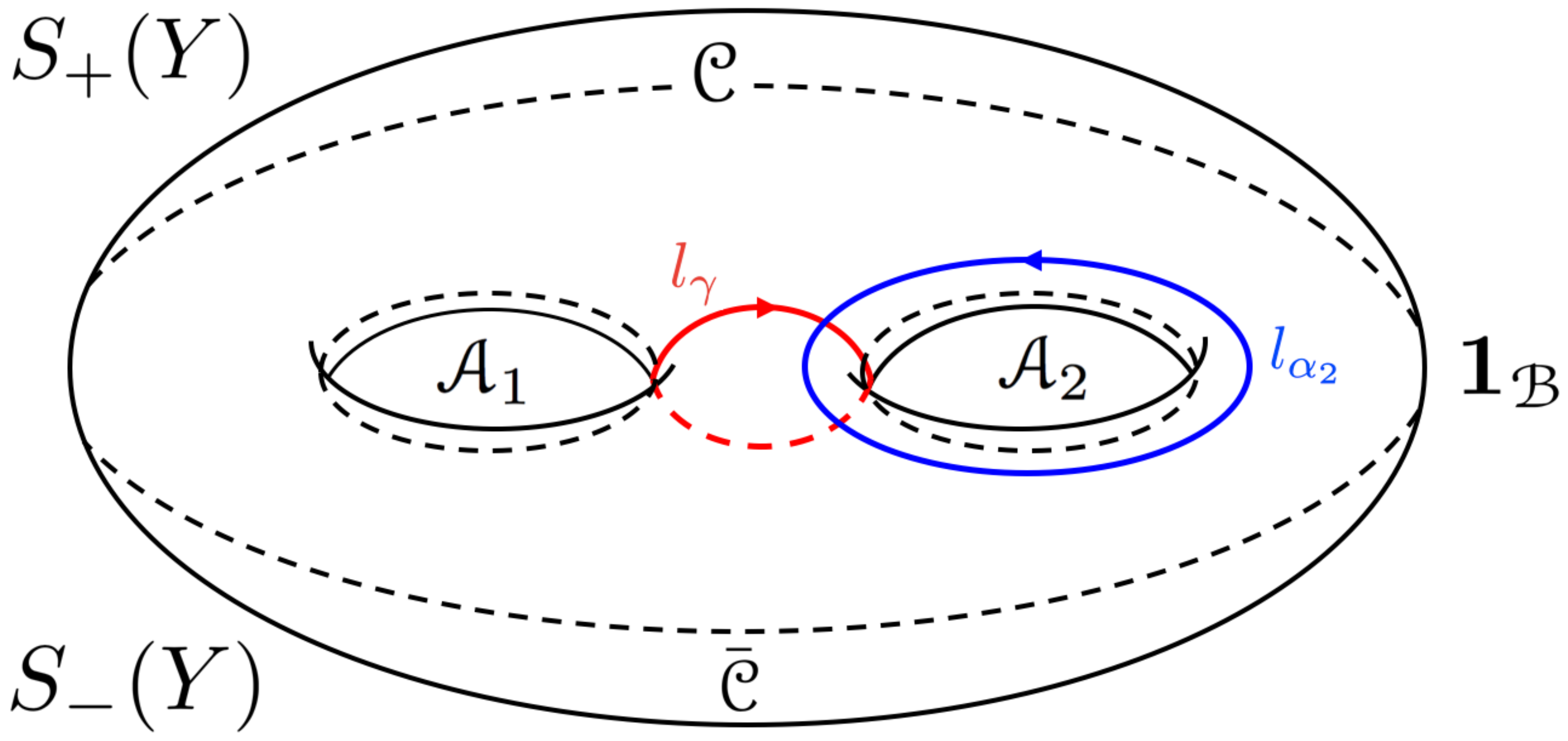}
\caption{Topological charge projection ($n$ = 2).}
\label{fig:tcp}
\end{figure}

One special case of TCM, namely topological charge projection \cite{Barkeshli16}, has been studied for doubled theories $\B$. In this case, $\B$ splits into two theories $\B = \mC \boxtimes \overbar{\mC}$ which do not interact in the bulk, but are ``stuck together'' at the original boundaries of $\B$. The planar region $Y$ also splits into two mirror layers, $S_{+}(Y)$ and $S_{-}(Y)$, which are completely disjoint in the bulk but ``stuck together'' at the boundaries of $Y$. Fig. \ref{fig:tcp} illustrates this for $n=2$. Notice that $S_+$ and $S_-$ together becomes a surface with genus $2$.

Consider the Wilson operator algebra $\mathcal{W}(\B,\{\A_i\})$ in this context. Each loop $\alpha_i \in \Gamma(Y)$ becomes a loop $l_{\alpha_i}$ in $S_{+}(Y)$ or $S_{-}(Y)$, while each arc $\gamma_i \in \Gamma(Y)$ lifts to a loop $l_{\gamma_i}$ going around both layers. Let $\beta$ be one of these loops. Define $\mathcal{O}_x(\beta)=W_x(\alpha_i)$ (tunneling operator in $\mC$) if $\beta$ is the lifting of the line $\alpha_i$, and $\mathcal{O}_x(\beta)=W_{x\overbar{x}}(\gamma_i)$ (loop operator in $\B$) if $\beta$ is the lifting of the loop $\gamma_i$. The Wilson operator measuring charge $a$ through $\beta$ gives the TCM \cite{Barkeshli16}
\begin{equation}
\label{eq:tcp}
P^{(a)}_{\beta}=\sum_{x\in \CC}S_{0a}S_{xa}^{*} \mathcal{O}_x(\beta). 
\end{equation}
\noindent
The sum runs over the anyon labels $x$ of $\mC$, and $S_{ab}$ is the modular $\mathcal{S}$-matrix of $\CC$. The Wilson operators $W_x(\alpha_i)$ and $W_{x\overbar{x}}(\gamma_i)$ are computed using the formulas (\ref{eq:tunnel-formula}) and (\ref{eq:loop-formula}) with the data of $\mC$ and $\B$, respectively.

\end{appendix}